\documentclass[twocolumn]{aastex631}

\usepackage{xcolor}
\usepackage{CJK}
\usepackage{threeparttable}
\usepackage[utf8]{inputenc}

\newcommand{\kms}{km\,s$^{-1}$}

\begin{document}
\title{Early Shock-Cooling Observations and Progenitor Constraints of Type IIb SN 2024uwq}

\correspondingauthor{B. Subrayan}
\email{bsubrayan@arizona.edu}

\newcommand{\LCO}{\affiliation{Las Cumbres Observatory, 6740 Cortona Drive, Suite 102, Goleta, CA 93117-5575, USA}}
\newcommand{\UCSB}{\affiliation{Department of Physics, University of California, Santa Barbara, CA 93106-9530, USA}}
\newcommand{\KITP}{\affiliation{Kavli Institute for Theoretical Physics, University of California, Santa Barbara, CA 93106-4030, USA}}
\newcommand{\UCD}{\affiliation{Department of Physics and Astronomy, University of California, Davis, 1 Shields Avenue, Davis, CA 95616-5270, USA}}
\newcommand{\WIS}{\affiliation{Department of Particle Physics and Astrophysics, Weizmann Institute of Science, 76100 Rehovot, Israel}}

\newcommand{\OKC}{\affiliation{Oskar Klein Centre, Department of Astronomy, Stockholm University, Albanova University Centre, SE-106 91 Stockholm, Sweden}}
\newcommand{\OAPD}{\affiliation{INAF-Osservatorio Astronomico di Padova, Vicolo dell'Osservatorio 5, I-35122 Padova, Italy}}
\newcommand{\Caltech}{\affiliation{Cahill Center for Astronomy and Astrophysics, California Institute of Technology, Mail Code 249-17, Pasadena, CA 91125, USA}}
\newcommand{\GSFC}{\affiliation{Astrophysics Science Division, NASA Goddard Space Flight Center, Mail Code 661, Greenbelt, MD 20771, USA}}
\newcommand{\UMD}{\affiliation{Joint Space-Science Institute, University of Maryland, College Park, MD 20742, USA}}
\newcommand{\UCB}{\affiliation{Department of Astronomy, University of California, Berkeley, CA 94720-3411, USA}}
\newcommand{\TTU}{\affiliation{Department of Physics, Texas Tech University, Box 41051, Lubbock, TX 79409-1051, USA}}
\newcommand{\STScI}{\affiliation{Space Telescope Science Institute, 3700 San Martin Drive, Baltimore, MD 21218-2410, USA}}
\newcommand{\UT}{\affiliation{University of Texas at Austin, 1 University Station C1400, Austin, TX 78712-0259, USA}}
\newcommand{\IoA}{\affiliation{Institute of Astronomy, University of Cambridge, Madingley Road, Cambridge CB3 0HA, UK}}
\newcommand{\QUB}{\affiliation{Astrophysics Research Centre, School of Mathematics and Physics, Queen's University Belfast, Belfast BT7 1NN, UK}}
\newcommand{\IPACSSC}{\affiliation{Spitzer Science Center, California Institute of Technology, Pasadena, CA 91125, USA}}
\newcommand{\IPAC}{\affiliation{IPAC, California Institute of Technology, 1200 East California Boulevard, Pasadena, CA 91125, USA}}
\newcommand{\JPL}{\affiliation{Jet Propulsion Laboratory, California Institute of Technology, 4800 Oak Grove Dr, Pasadena, CA 91109, USA}}
\newcommand{\Southampton}{\affiliation{Department of Physics and Astronomy, University of Southampton, Southampton SO17 1BJ, UK}}
\newcommand{\LANL}{\affiliation{Space and Remote Sensing, MS B244, Los Alamos National Laboratory, Los Alamos, NM 87545, USA}}
\newcommand{\Tsinghua}{\affiliation{Physics Department and Tsinghua Center for Astrophysics, Tsinghua University, Beijing, 100084, People's Republic of China}}
\newcommand{\NAOC}{\affiliation{National Astronomical Observatory of China, Chinese Academy of Sciences, Beijing, 100012, People's Republic of China}}
\newcommand{\YNAO}{\affiliation{Yunnan Observatories (YNAO), Chinese Academy of Sciences (CAS), Kunming, 650216, People's Republic of China}}
\newcommand{\ICEY}{\affiliation{International Centre of Supernovae, Yunnan Key Laboratory, Kunming 650216, People's Republic of China}}
\newcommand{\Itagaki}{\affiliation{Itagaki Astronomical Observatory, Yamagata 990-2492, Japan}}
\newcommand{\Einstein}{\altaffiliation{Einstein Fellow}}
\newcommand{\Hubble}{\altaffiliation{Hubble Fellow}}
\newcommand{\CfA}{\affiliation{Center for Astrophysics \textbar{} Harvard \& Smithsonian, 60 Garden Street, Cambridge, MA 02138-1516, USA}}
\newcommand{\UA}{\affiliation{Steward Observatory, University of Arizona, 933 North Cherry Avenue, Tucson, AZ 85721-0065, USA}}
\newcommand{\MPIA}{\affiliation{Max-Planck-Institut f\"ur Astrophysik, Karl-Schwarzschild-Stra\ss{}e 1, D-85748 Garching, Germany}}
\newcommand{\DSFP}{\altaffiliation{LSSTC Data Science Fellow}}
\newcommand{\HCO}{\affiliation{Harvard College Observatory, 60 Garden Street, Cambridge, MA 02138-1516, USA}}
\newcommand{\Carnegie}{\affiliation{Observatories of the Carnegie Institute for Science, 813 Santa Barbara Street, Pasadena, CA 91101-1232, USA}}
\newcommand{\TAU}{\affiliation{School of Physics and Astronomy, Tel Aviv University, Tel Aviv 69978, Israel}}
\newcommand{\Edinburgh}{\affiliation{Institute for Astronomy, University of Edinburgh, Royal Observatory, Blackford Hill EH9 3HJ, UK}}
\newcommand{\Birmingham}{\affiliation{Birmingham Institute for Gravitational Wave Astronomy and School of Physics and Astronomy, University of Birmingham, Birmingham B15 2TT, UK}}
\newcommand{\Bath}{\affiliation{Department of Physics, University of Bath, Claverton Down, Bath BA2 7AY, UK}}
\newcommand{\CTIO}{\affiliation{Cerro Tololo Inter-American Observatory, National Optical Astronomy Observatory, Casilla 603, La Serena, Chile}}
\newcommand{\Potsdam}{\affiliation{Institut f\"ur Physik und Astronomie, Universit\"at Potsdam, Haus 28, Karl-Liebknecht-Str. 24/25, D-14476 Potsdam-Golm, Germany}}
\newcommand{\INPE}{\affiliation{Instituto Nacional de Pesquisas Espaciais, Avenida dos Astronautas 1758, 12227-010, S\~ao Jos\'e dos Campos -- SP, Brazil}}
\newcommand{\UNC}{\affiliation{Department of Physics and Astronomy, University of North Carolina, 120 East Cameron Avenue, Chapel Hill, NC 27599, USA}}
\newcommand{\Ohio}{\affiliation{Astrophysical Institute, Department of Physics and Astronomy, 251B Clippinger Lab, Ohio University, Athens, OH 45701-2942, USA}}
\newcommand{\AAS}{\affiliation{American Astronomical Society, 1667 K~Street NW, Suite 800, Washington, DC 20006-1681, USA}}
\newcommand{\MMT}{\affiliation{MMT and Steward Observatories, University of Arizona, 933 North Cherry Avenue, Tucson, AZ 85721-0065, USA}}
\newcommand{\Geneva}{\affiliation{ISDC, Department of Astronomy, University of Geneva, Chemin d'\'Ecogia, 16 CH-1290 Versoix, Switzerland}}
\newcommand{\IUCAA}{\affiliation{Inter-University Center for Astronomy and Astrophysics, Post Bag 4, Ganeshkhind, Pune, Maharashtra 411007, India}}
\newcommand{\CMU}{\affiliation{Department of Physics, Carnegie Mellon University, 5000 Forbes Avenue, Pittsburgh, PA 15213-3815, USA}}
\newcommand{\NAOJ}{\affiliation{Division of Science, National Astronomical Observatory of Japan, 2-21-1 Osawa, Mitaka, Tokyo 181-8588, Japan}}
\newcommand{\IfA}{\affiliation{Institute for Astronomy, University of Hawai`i, 2680 Woodlawn Drive, Honolulu, HI 96822-1839, USA}}
\newcommand{\UCSC}{\affiliation{Department of Astronomy and Astrophysics, University of California, Santa Cruz, CA 95064-1077, USA}}
\newcommand{\Purdue}{\affiliation{Department of Physics and Astronomy, Purdue University, 525 Northwestern Avenue, West Lafayette, IN 47907-2036, USA}}
\newcommand{\Princeton}{\affiliation{Department of Astrophysical Sciences, Princeton University, 4 Ivy Lane, Princeton, NJ 08540-7219, USA}}
\newcommand{\Moore}{\affiliation{Gordon and Betty Moore Foundation, 1661 Page Mill Road, Palo Alto, CA 94304-1209, USA}}
\newcommand{\Durham}{\affiliation{Department of Physics, Durham University, South Road, Durham, DH1 3LE, UK}}
\newcommand{\JHU}{\affiliation{Department of Physics and Astronomy, The Johns Hopkins University, 3400 North Charles Street, Baltimore, MD 21218, USA}}
\newcommand{\Toronto}{\affiliation{David A.\ Dunlap Department of Astronomy and Astrophysics, University of Toronto,\\ 50 St.\ George Street, Toronto, Ontario, M5S 3H4 Canada}}
\newcommand{\Duke}{\affiliation{Department of Physics, Duke University, Campus Box 90305, Durham, NC 27708, USA}}
\newcommand{\NCU}{\affiliation{Graduate Institute of Astronomy, National Central University, 300 Jhongda Road, 32001 Jhongli, Taiwan}}
\newcommand{\Columbia}{\affiliation{Department of Physics and Columbia Astrophysics Laboratory, Columbia University, Pupin Hall, New York, NY 10027, USA}}
\newcommand{\Flatiron}{\affiliation{Center for Computational Astrophysics, Flatiron Institute, 162 5th Avenue, New York, NY 10010-5902, USA}}
\newcommand{\CIERA}{\affiliation{Center for Interdisciplinary Exploration and Research in Astrophysics and Department of Physics and Astronomy, \\Northwestern University, 1800 Sherman Avenue, 8th Floor, Evanston, IL 60201, USA}}
\newcommand{\GeminiNorth}{\affiliation{Gemini Observatory, 670 North A`ohoku Place, Hilo, HI 96720-2700, USA}}
\newcommand{\SNU}{\affiliation{Department of Physics and Astronomy, Seoul National University, Gwanak-ro 1, Gwanak-gu, Seoul, 08826, South Korea}}

\newcommand{\GeminiNOIRLab}{\affiliation{Gemini Observatory/NSF's National Optical-Infrared Astronomy Research Laboratory, 670 N. Aohoku Place, Hilo, HI, 96720, USA;tom.geballe\@noirlab.edu}}

\newcommand{\Keck}{\affiliation{W.~M.~Keck Observatory, 65-1120 M\=amalahoa Highway, Kamuela, HI 96743-8431, USA}}
\newcommand{\UW}{\affiliation{Department of Astronomy, University of Washington, 3910 15th Avenue NE, Seattle, WA 98195-0002, USA}}
\newcommand{\catalyst}{\altaffiliation{LSSTC Catalyst Fellow}}
\newcommand{\USask}{\affiliation{Department of Physics \& Engineering Physics, University of Saskatchewan, 116 Science Place, Saskatoon, SK S7N 5E2, Canada}}
\newcommand{\Thacher}{\affiliation{Thacher School, 5025 Thacher Road, Ojai, CA 93023-8304, USA}}
\newcommand{\Rutgers}{\affiliation{Department of Physics and Astronomy, Rutgers, the State University of New Jersey,\\136 Frelinghuysen Road, Piscataway, NJ 08854-8019, USA}}
\newcommand{\FSU}{\affiliation{Department of Physics, Florida State University, 77 Chieftan Way, Tallahassee, FL 32306-4350, USA}}
\newcommand{\Melbourne}{\affiliation{School of Physics, The University of Melbourne, Parkville, VIC 3010, Australia}}
\newcommand{\ASTROthreeD}{\affiliation{ARC Centre of Excellence for All Sky Astrophysics in 3 Dimensions (ASTRO 3D)}}
\newcommand{\Stromlo}{\affiliation{Mt.\ Stromlo Observatory, The Research School of Astronomy and Astrophysics, Australian National University, ACT 2601, Australia}}
\newcommand{\NCPAS}{\affiliation{National Centre for the Public Awareness of Science, Australian National University, Canberra, ACT 2611, Australia}}
\newcommand{\TAMU}{\affiliation{Department of Physics and Astronomy, Texas A\&M University, 4242 TAMU, College Station, TX 77843, USA}}
\newcommand{\Mitchell}{\affiliation{George P.\ and Cynthia Woods Mitchell Institute for Fundamental Physics \& Astronomy, College Station, TX 77843, USA}}
\newcommand{\ESO}{\affiliation{European Southern Observatory, Alonso de C\'ordova 3107, Casilla 19, Santiago, Chile}}
\newcommand{\ICE}{\affiliation{Institute of Space Sciences (ICE, CSIC), Campus UAB, Carrer
de Can Magrans, s/n, E-08193 Barcelona, Spain}}
\newcommand{\IEEC}{\affiliation{Institut d'Estudis Espacials de Catalunya (IEEC), Edifici RDIT, Campus UPC, 08860 Castelldefels (Barcelona), Spain}}
\newcommand{\Warwick}{\affiliation{Department of Physics, University of Warwick, Gibbet Hill Road, Coventry CV4 7AL, UK}}
\newcommand{\Macquarie}{\affiliation{School of Mathematical and Physical Sciences, Macquarie University, NSW 2109, Australia}}
\newcommand{\AAARC}{\affiliation{Astronomy, Astrophysics and Astrophotonics Research Centre, Macquarie University, Sydney, NSW 2109, Australia}}
\newcommand{\Capodimonte}{\affiliation{INAF - Capodimonte Astronomical Observatory, Salita Moiariello 16, I-80131 Napoli, Italy}}
\newcommand{\INFNNapoli}{\affiliation{INFN - Napoli, Strada Comunale Cinthia, I-80126 Napoli, Italy}}
\newcommand{\ICRANet}{\affiliation{ICRANet, Piazza della Repubblica 10, I-65122 Pescara, Italy}}
\newcommand{\MSU}{\affiliation{Center for Data Intensive and Time Domain Astronomy, Department of Physics and Astronomy,\\Michigan State University, East Lansing, MI 48824, USA}}
\newcommand{\SETI}{\affiliation{SETI Institute,
339 Bernardo Ave, Suite 200, Mountain View, CA 94043, USA}} 
\newcommand{\IAIFI}{\affiliation{The NSF AI Institute for Artificial Intelligence and Fundamental Interactions}}
\newcommand{\ANUC}{\affiliation{Department of Astronomy, AlbaNova University Center, Stockholm University, SE-10691 Stockholm, Sweden}}

\newcommand{\Konkoly}{\affiliation{Konkoly Observatory,  CSFK, Konkoly-Thege M. \'ut 15-17, Budapest, 1121, Hungary}}
\newcommand{\ELTE}{\affiliation{ELTE E\"otv\"os Lor\'and University, Institute of Physics, P\'azm\'any P\'eter s\'et\'any 1/A, Budapest, 1117 Hungary}}
\newcommand{\SZTE}{\affiliation{Department of Experimental Physics, University of Szeged, D\'om t\'er 9, Szeged, 6720, Hungary}}
\newcommand{\IdAlta}{\affiliation{Instituto de Alta Investigaci\'on, Sede Esmeralda, Universidad de Tarapac\'a, Av. Luis Emilio Recabarren 2477, Iquique, Chile}}
\newcommand{\Kavli}{\affiliation{Kavli Institute for Cosmological Physics, University of Chicago, Chicago, IL 60637, USA}}
\newcommand{\UofChicago}{\affiliation{Department of Astronomy and Astrophysics, University of Chicago, Chicago, IL 60637, USA}}
\newcommand{\Fermi}{\affiliation{Fermi National Accelerator Laboratory, P.O.\ Box 500, Batavia, IL 60510, USA}}
\newcommand{\Dartmouth}{\affiliation{Department of Physics and Astronomy, Dartmouth College, Hanover, NH 03755, USA}}
\newcommand{\Surrey}{\affiliation{Department of Physics, University of Surrey, Guildford GU2 7XH, UK}}
\newcommand{\NU}{\affiliation{Center for Interdisciplinary Exploration and Research in Astrophysics (CIERA), Northwestern University, Evanston, IL 60208, USA}}
\newcommand{\itagaki}{\affiliation{Itagaki Astronomical Observatory, Yamagata 990-2492, Japan}}
\newcommand{\UdChile}{\affiliation{Departamento de Astronomia, Universidad de Chile, Camino El Observatorio 1515, Las Condes, Santiago, Chile}}
\newcommand{\UVA}{\affiliation{Department of Astronomy, University of Virginia, Charlottesville, VA 22904, USA}}
\newcommand{\UCSD}{\affiliation{Department of Astronomy \& Astrophysics, University of California, San Diego, 9500 Gilman Drive, MC 0424, La Jolla, CA 92093-0424, USA}}

\author[0000-0001-8073-8731]{Bhagya M.\ Subrayan}
\UA

\author[0000-0003-4102-380X]{David J. Sand}
\UA

\author[0000-0002-4924-444X]{K. Azalee Bostroem}
\catalyst\UA

\author[0000-0001-8738-6011]{Saurabh W.\ Jha}
\Rutgers

\author[0000-0002-7352-7845]{Aravind P.\ Ravi}
\UCD

\author[0009-0002-5096-1689]{Michaela Schwab}
\Rutgers

\author[0000-0003-0123-0062]{Jennifer E. Andrews}
\GeminiNorth

\author[0000-0002-0832-2974]{Griffin Hosseinzadeh}
\UCSD

\author[0000-0001-8818-0795]{Stefano Valenti}
\UCD

\author[0000-0002-7937-6371]{Yize Dong \begin{CJK*}{UTF8}{gbsn}(董一泽)\end{CJK*}}
\UCD

\author[0000-0002-0744-0047]{Jeniveve Pearson}
\UA

\author[0000-0002-4022-1874]{Manisha Shrestha}
\UA

\author[0000-0003-3108-1328]{Lindsey~A.~Kwok}
\NU

\author[0000-0003-2744-4755]{Emily Hoang}
\UCD

\author[0000-0003-3643-839X]{Jeonghee Rho}
\SETI

\author[0000-0001-7488-4337]{Seong Hyun Park}
\SNU
\author[0000-0002-5847-8096]{Sung-Chul Yoon}
\SNU
\author[0000-0003-2824-3875]{T. R. Geballe}
\GeminiNorth
\author[0000-0002-6703-805X]{Joshua Haislip}
\UNC

\author[0000-0003-0549-3281]{Daryl Janzen}
\USask

\author[0000-0003-3642-5484]{Vladimir Kouprianov}
\UNC



\author[0009-0008-9693-4348]{Darshana Mehta}
\UCD

\author[0000-0002-7015-3446]{Nicol\'as Meza Retamal}
\UCD

\author[0000-0002-5060-3673]{Daniel E.\ Reichart}
\UNC



\author[0000-0002-1895-6639
]{Moira Andrews}
\LCO\UCSB

\author[0000-0003-4914-5625]{Joseph Farah}
\LCO 
\UCSB

\author[0000-0001-9570-0584]{Megan Newsome}
\LCO 
\UCSB

\author[0000-0003-4253-656X]{D.\ Andrew Howell}
\LCO\UCSB
\author[0000-0001-5807-7893]{Curtis McCully}
\LCO
















\begin{abstract}
We present early multi-wavelength photometric and spectroscopic observations of the Type IIb supernova SN 2024uwq, capturing its shock-cooling emission phase and double-peaked light curve evolution. Early spectra reveal broad H$\alpha$ ($v \sim 15,500$ km s$^{-1}$) and He I P-Cygni profiles of similar strengths. Over time the He I lines increase in strength while the H$\alpha$ decreases, consistent with a hydrogen envelope ($M_{\text{env}} = 0.7$–1.35\,$M_\odot$) overlying helium-rich ejecta. Analytic modeling of early shock cooling emission and bolometric light analysis constrains the progenitor to a partially stripped star with radius $R = 10 - 60\,R_\odot$, consistent with a blue/yellow supergiant with an initial ZAMS mass of 12–20\,$M_\odot$ likely stripped via binary interaction. SN~2024uwq occupies a transitional position between compact and extended Type IIb supernovae, highlighting the role of binary mass-transfer efficiency in shaping a continuum of stripped-envelope progenitors. Our results underscore the importance of both early UV/optical observations to characterize shock breakout signatures critical to map the diversity in evolutionary pathways of massive stars. Upcoming time domain surveys including Rubin Observatory’s LSST and UV missions like \textit{ULTRASAT} and \textit{UVEX} will revolutionize our ability to systematically capture these early signatures, probing the full diversity of stripped progenitors and their explosive endpoints.
\end{abstract}

\defcitealias{SW2017}{SW17}
\defcitealias{Piro2015}{P15}
\defcitealias{Piro2021}{P21}
\defcitealias{MSW2023}{MSW23}

\section{Introduction} \label{sec:intro}

Massive stars (\(\gtrsim 8\,M_{\odot}\)) explode as core collapse supernovae (CCSNe). While the majority of CCSNe show hydrogen in their spectra, a subset undergoes extensive mass loss, shedding their outer H and He layers to become stripped envelope supernovae (SESNe; \citealt{Woosley1994a,F1997, G2017}). Among CCSNe, Type IIb supernovae are transitional objects - while their early spectra have weak hydrogen lines, these lines fade within weeks, revealing helium dominated profiles similar to those of SNe Ib - hydrogen poor explosions marked by strong helium lines in their optical spectra (e.g., SN~1987K: \citealt{F1988}; SN~1993J: \citealt{F1993, R1994}; SN~2008ax: \citealt{Pastorello2008}; SN~2011dh: \citealt{A2011, Soderberg2012}; SN~2011ei: \citealt{Dan2013}; SN~2011fu: \citealt{Morales2015_2011fu}; SN~2013df: \citealt{V2014, morales2014_2013df}; SN~2016gkg: \citealt{Arcavi2017, Tartagalia2017, Bersten2018}). This spectral evolution indicates that the progenitors retain only a thin hydrogen envelope (\(\lesssim 1\,M_{\odot}\)) at explosion, offering a unique window into the final stages of massive star evolution \citep{R1994, Mathesan2000}.  Although it critically shapes the final structure of the progenitor, this extensive mass loss driven by mechanisms such as stellar winds \citep{Woos1993,Groh2013,Geor2013}, binary interactions \citep{P2008,Smith2014, O2017, S2017}, or rotational stripping \citep{G2013} remains poorly understood. Observational studies reveal a diverse progenitor population, including yellow supergiants (e.g., SN 2011dh), K-supergiants (e.g., SN 1993J) and Wolf-Rayet stars (e.g., SN 2008ax), spanning initial masses of 10--28 $M_{\odot}$ \citep{F1993,A2011,C2008}. 
Probing circumstellar material (CSM) from the radio/X-ray counterparts of these SNe has also provided constraints on the wind velocities and mass-loss history responsible for the stripping of the outer envelope. Such diversity underscores the complex interplay of binary evolution and stellar physics in shaping pre-supernova systems \citep{S2020}.

A characteristic of many SNe IIb's is their double-peaked light curve. The brief initial peak, lasting hours to days, arises from shock-cooling emission (SCE) as the explosion's thermalized energy radiates from the extended envelope of the progenitor \citep{R1994, Arcavi2017, Das2023}. Analytical and numerical models \citep{Rabinak2011,Piro2015,Piro2021} link the cooling phase with the density and radius of the envelope with recent extensions incorporating multi zone dynamics and UV line blanketing \citep{SW2017, MSW2023}.
These models, when applied to high cadence observations of nearby SNe IIb which resolve the SCE phase \citep{R1994, Arcavi2017, Armstrong_2017jgh,F2025_2022hnt}, suggest progenitors with extended envelopes (\(\sim 100\)-\(500\,R_{\odot}\)) and low residual hydrogen masses (\(\sim 0.01\)-\(1\,M_{\odot}\)), consistent with pre-explosion imaging of yellow and red supergiants \citep{Tartagalia2017,Bersten2018, Kilpatrick2022}. 

In this Letter, we present a comprehensive analysis of SN~2024uwq, a nearby (\textit{D} $\approx$ 47 Mpc) SN IIb with early multi-wavelength photometric and spectroscopic observations.  In Section \ref{sec:discovery}, we detail the discovery, distance estimation, and reddening considerations for SN 2024uwq. Section \ref{sec:obs} describes our observations and data reduction procedures, which include both imaging and optical spectroscopy. In Section \ref{sec:lc_analysis}, we analyze the photometric data, focusing on the early shock-cooling emission, color evolution, bolometric luminosity, and estimates of the synthesized $^{56}$Ni mass. Section \ref{sec:spectra} presents the spectroscopic features and their temporal evolution, comparing them with other Type IIb supernovae. We present early shock-cooling emission modeling to constrain progenitor properties for various analytical frameworks in Section \ref{sec:sc_model}. In Section \ref{sec:results}, we report our results and findings on SN~2024uwq, situating it within the broader context of SESNe and discussing its implications for progenitor scenarios. Section \ref{sec:conclusion} then summarizes our conclusions and outlines the prospects for early, high-cadence follow-up observations with upcoming missions such as \textit{ULTRASAT}, \textit{UVEX}, and LSST.

\begin{figure}
    \centering
    \includegraphics[width=\linewidth]{ 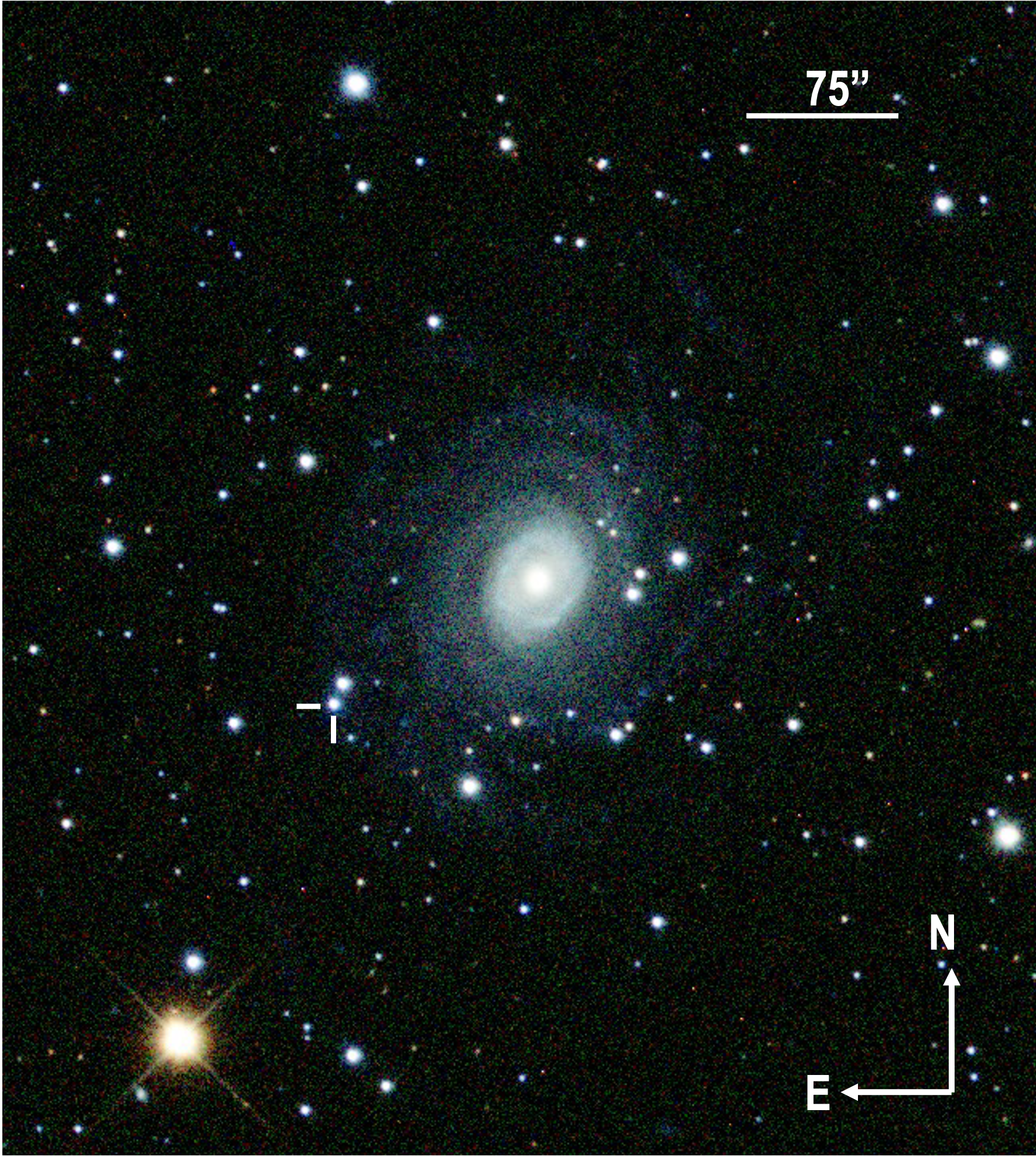}
    \caption{Composite \textit{gri} image of SN~2024uwq obtained using Las Cumbres Observatory observations on September 28, 2024. SN~2024uwq is marked with white cross hairs in the composite image. }
    \label{fig:SN2024uwq_field}
    \vspace{-0.5 em}
\end{figure}

\section{Discovery, Distance and Reddening}\label{sec:discovery}

SN 2024uwq was discovered by the Asteroid Terrestrial-impact Last Alert System  (ATLAS; \citealt{Tonry2018,Smith2020}) on 2024-09-07 02:33:57.02 UT (MJD = 60560.11) with a discovery magnitude of $o = 17.38$ mag \citep{Tonry2024}. All dates and times used in this work are reported in the Coordinated Universal Time (UTC) standard. SN 2024uwq is located in NGC 6902 (Figure \ref{fig:SN2024uwq_field}) at J2000 coordinates $\alpha = 20^{\text{h}}24^{\text{m}}36.770^{\text{s}}$ and $\delta = -43^{\circ}40^{\arcmin}10.13^{\arcsec}$. The last available non-detection was reported by ATLAS on 2024-09-04 at 03:51:10 (MJD = 60557.16), $\sim$ 3 days prior to the discovery date, with a limiting magnitude of $o = 19.3$ mag. Throughout this paper, we adopt the explosion date (t$_0$) to be the midpoint between the last ATLAS non-detection and discovery date which is at MJD = 60558.63 $\pm 1.5$, where the uncertainty covers the time between non-detection and discovery. Unless stated otherwise, all phases reported in this work are calculated using this explosion date.

SN 2024uwq was initially classified as a SN Ic-BL by the extended Public European Southern Observatory (ESO) Spectroscopic Survey of Transient Objects (ePESSTO+: \citealt{Smartt2015}) at a redshift $z = 0.009$ \citep{Ramirez2024}. On 2024-09-17, SN 2024uwq was reclassified as a Type IIb supernova using a spectrum taken by the Global Supernova Project (GSP; \citet{Howell2024}). This classification, based on GELATO (\citealt{H2008}) and Supernova Identification (SNID: \citealt{Blondin2011}) code comparisons, shows the spectrum best matches young Type IIb supernova templates with redshifts between 0.003 - 0.009 \citep{Bostroem2024}. We adopted a redshift of $z=0.009$ in this work, as this value aligns closely with supernova templates and is confirmed by the Na~I~D absorption features detected in our highest signal-to-noise spectra. We use the Tully-Fisher \citep{Tully2009} distance modulus value of $\mu = 33.34\,\pm 0.40$ mag that yields a distance of $D = 46.6\,\pm 8.6$ Mpc, which is adopted throughout this paper. 

\begin{table}
 \caption{Properties of SN~2024uwq} \label{tab:specs}
 \begin{tabular}{ll}
    \hline
    Parameter & Value \\
    \hline
    R.A. (J2000) & 20:24:36.76 \\
    Dec. (J2000) & $-$43:40:09.9\\
    Last Non-detection (MJD) & 60557.16  \\ 
    First Detection (MJD) & 60560.10 \\
    Explosion Epoch (MJD)\tablenotemark{a} & $60558.63 \pm 1.5$ \\
    Redshift ($z$)\tablenotemark{b} & 0.009  \\
    Distance modulus \tablenotemark{c}& 33.34 $\pm$ 0.40 mag\\
    Distance\tablenotemark{c} &  46.6 $\pm 8.6$ Mpc\\
    $E(B-V)_\mathrm{MW}$ & $0.034 \pm 0.025$ mag\\
    $E(B-V)_\mathrm{host}$\tablenotemark{d} & $< 0.02$ mag\\
    $E(B-V)_\mathrm{total}$ \tablenotemark{e} & $0.034 \pm 0.001$ mag\\
    Peak Magnitude ($V_{\mathrm{max}}$) &  $-17.79 \pm 0.4$ mag \\
    \hline
 \end{tabular} 
 \tablenotetext{a}{mid point of last non-detection and first detection}
 \tablenotetext{b}{from best match SNID templates}
 \tablenotetext{c}{estimates from \citet{Tully2009}}
 \tablenotetext{d}{from the \ion{Na}{1}~D lines of the host galaxy}
\tablenotetext{e}{from \citet{Schlafly2011} MW dust maps}
\end{table}

To estimate reddening along the line of sight to SN~2024uwq, we considered contributions from both the Milky Way (MW) and the host galaxy. Using the high signal-to-noise ratio (SNR) SALT spectrum ($R \sim 600-2000$) obtained on 2024 October 17, we measured the equivalent widths (EWs) of Na~\textsc{i}~D absorption lines, which are empirically correlated with reddening due to their association with interstellar gas and dust \citep{Poznanski2012}. We continuum-normalized the observed spectrum and modeled the blended Na~\textsc{i}~D$_2$ and Na~\textsc{i}~D$_1$ absorption lines from MW with a single Gaussian profile. This yielded a total EW of $0.32 \pm 0.04$~\AA. Using the relationship between EW and reddening as given in \citet{Poznanski2012}, we derived a Milky Way reddening of $E(B-V)_{\text{MW}} = 0.034 \pm 0.025$~mag. We compared our above  reddening estimate with the dust maps of \citet{Schlafly2011}, which give $E(B-V)_{\text{MW}} = 0.034 \pm 0.001$~mag for the direction of SN~2024uwq. This value is consistent with our Na~\textsc{i}~D-based measurement.

For the host galaxy, we inspected the observed spectrum for Na~\textsc{i}~D absorption features at observed wavelengths D$_2$ ($\lambda 5949$) and D$_1$ ($\lambda 5943$) corresponding to the rest frame D$_2$ ($\lambda 5890$) and D$_1$ ($\lambda 5896$) lines. No significant absorption dips were detected, and we set an upper limit on the host galaxy EW of Na~\textsc{i}~D to $< 0.03$~\AA\ by measuring a 3$\sigma$ noise level in the continuum. This corresponds to a reddening of $E(B-V)_{\text{host}} < 0.02$~mag. Given that this upper limit is comparable to the uncertainty in $E(B-V)_{\text{MW}}$, we assume that the host galaxy's contribution to reddening is negligible. Therefore, we adopt the total reddening value $E(B-V)_{\text{total}} \approx 0.034 \pm 0.001$~mag, and apply the extinction law of \citet{Cardelli1989} with $R_V = 3.1$. Table \ref{tab:specs} summarizes the relevant physical quantities for SN~2024uwq.

\section{Observations and Data Reduction} \label{sec:obs}

\subsection{Imaging}

An extensive photometric campaign was launched immediately after the discovery of SN~2024uwq to ensure comprehensive coverage of its early light curve evolution. High cadence observations of SN~2024uwq  were performed in \textit{U, B, V, g, r, i} bands using the worldwide network of 0.4-m and 1-m telescopes available through the Las Cumbres Observatory with the Global Supernova Project \citep{Brown_2013}. Data were processed with the PyRAF-based pipeline \texttt{lcogtsnpipe} \citep{Valenti_2016} using PSF fitting. The \textit{UBV} magnitudes were calibrated in the Vega system against standard fields observed with the same telescope on the same night, using the Landolt catalog \citep{Landolt_1992}. For the \textit{gri} bands, calibrations were performed in the AB magnitude system using reference stars from the American Association of Variable Star Observers (AAVSO) Photometric All-Sky Survey (APASS; \citealt{Henden_2009}). The SN's significant spatial offset from the core regions of the host galaxy (see Figure \ref{fig:SN2024uwq_field}) resulted in minimal contamination, allowing direct PSF photometry on the images.

Additional early high-cadence photometry of SN~2024uwq was obtained as part of the Distance Less Than 40 Mpc (DLT40) survey \citep{Tartaglia_2018} using the PROMPT-MO  0.4-m telescope at Meckering Observatory in Australia, through the Skynet Robotic Telescope Network \citep{Reichart_2005}. Observations were conducted in \textit{B, V, g, r,} and \textit{i} bands, as well as in a filterless ``Open" wide band mode. The wide band data were calibrated to the SDSS \textit{r} band following the reduction procedures detailed in \citet{Tartaglia_2018}, while the multi-band aperture photometry, performed with \texttt{photutils} \citep{Bradley_2022}, were calibrated using the APASS catalog. 

All publicly available ATLAS photometry of SN~2024uwq observed in $c$ and $o$ bands were retrieved using the ATLAS forced photometry service \citep{Tonry2018, Smith2020} \footnote{\url{https://fallingstar-data.com/forcedphot/}}. 

\begin{figure*}
    \centering
    \includegraphics[width=\textwidth]{ 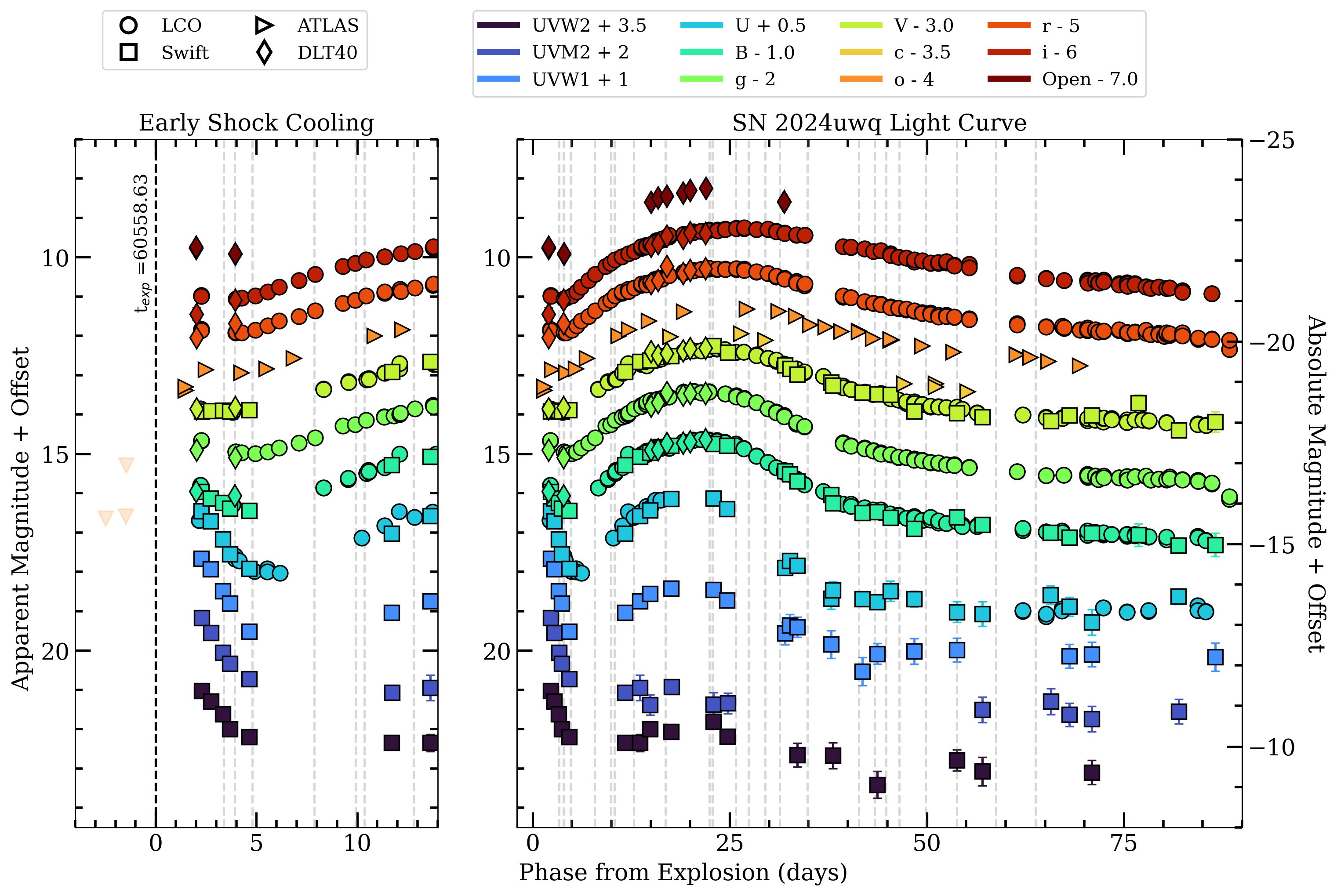}
    \caption{Multiwavelength observations of SN 2024uwq with early phases of the light curve showing characteristic shock cooling emission from the progenitor. The offsets for each bands are marked in the legend. The time of explosion is marked on the left panel, which zooms in the early light curve evolution. The observations provided in this figure are not corrected for extinction. The grey vertical lines mark the phases where optical spectra were obtained. }
    \label{fig:lc_early}
\end{figure*}

High cadence ultraviolet (UV) and optical observations of SN~2024uwq were also obtained with the Ultraviolet/Optical Telescope (UVOT; \citealt{Roming_2005}) onboard the \textit{Neil Gehrels Swift Observatory} (\citealt{Gehrels_2004}). The data, retrieved from the NASA \textit{Swift} Data Archive\footnote{\url{https://heasarc.gsfc.nasa.gov/cgi-bin/W3Browse/swift.pl}}, were processed using standard tools provided within the High-Energy Astrophysics software (\texttt{HEASoft}\footnote{\url{https://heasarc.gsfc.nasa.gov/docs/software/heasoft/}}) package. Photometry was performed in the \textit{uvw1}, \textit{uvm2}, \textit{uvw2}, \textit{U$_{S}$}, \textit{B$_{S}$}, and \textit{V$_{S}$} bands. A source aperture of 3\arcsec\ was used, centered on the supernova position, with background subtraction performed from nearby regions free of contaminating sources. Although no pre-explosion template images were available, host galaxy contamination was minimal and therefore not subtracted. The zero points for photometric calibration were adopted from \citet{Breeveld_2010}, incorporating time-dependent sensitivity corrections updated in 2020. All light curves derived from imaging observations are presented in Figure \ref{fig:lc_early}.

\begin{figure*}
    \centering
    \includegraphics[width=0.45\textwidth]{ 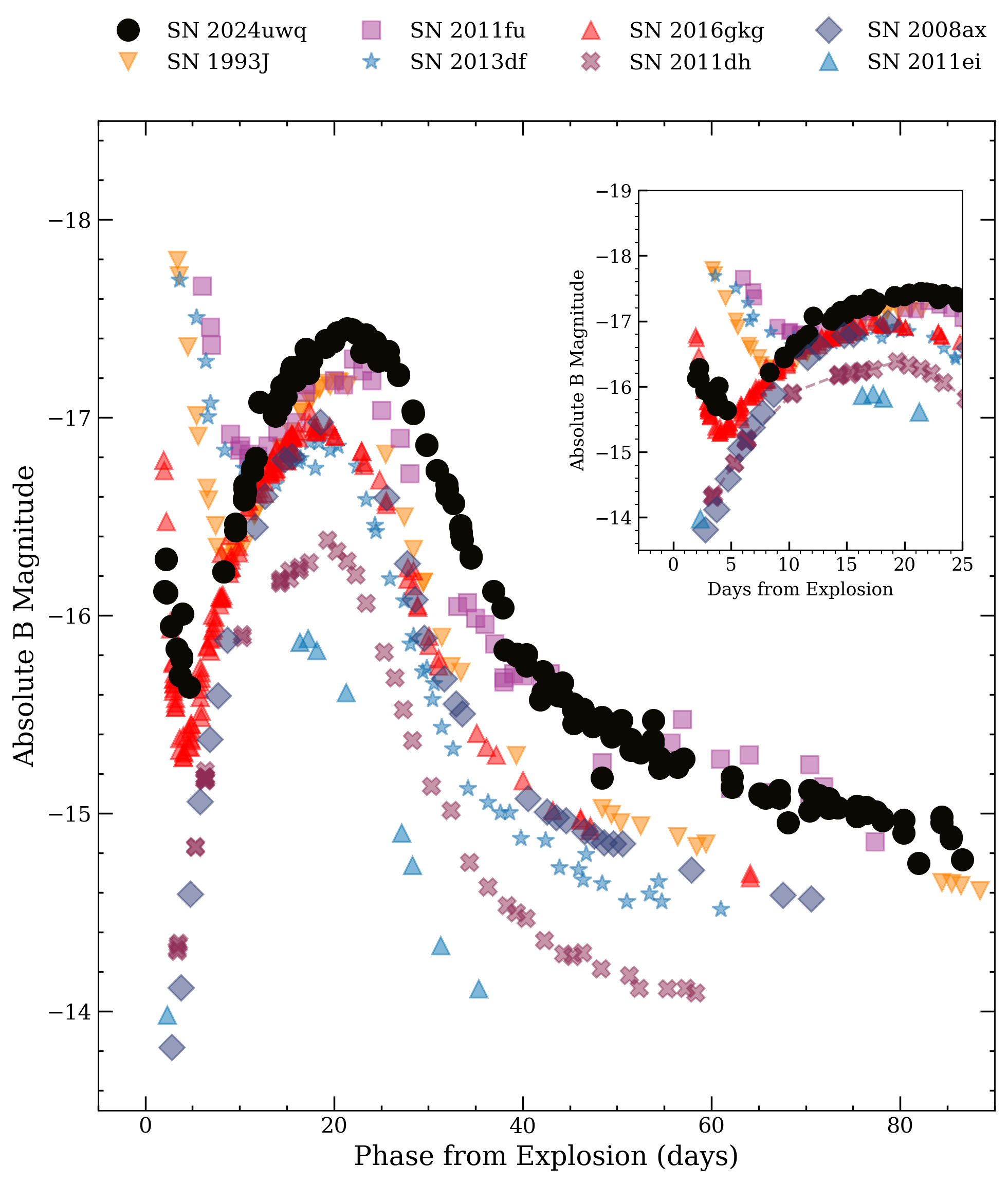}
    \includegraphics[width =0.45 \textwidth]{ 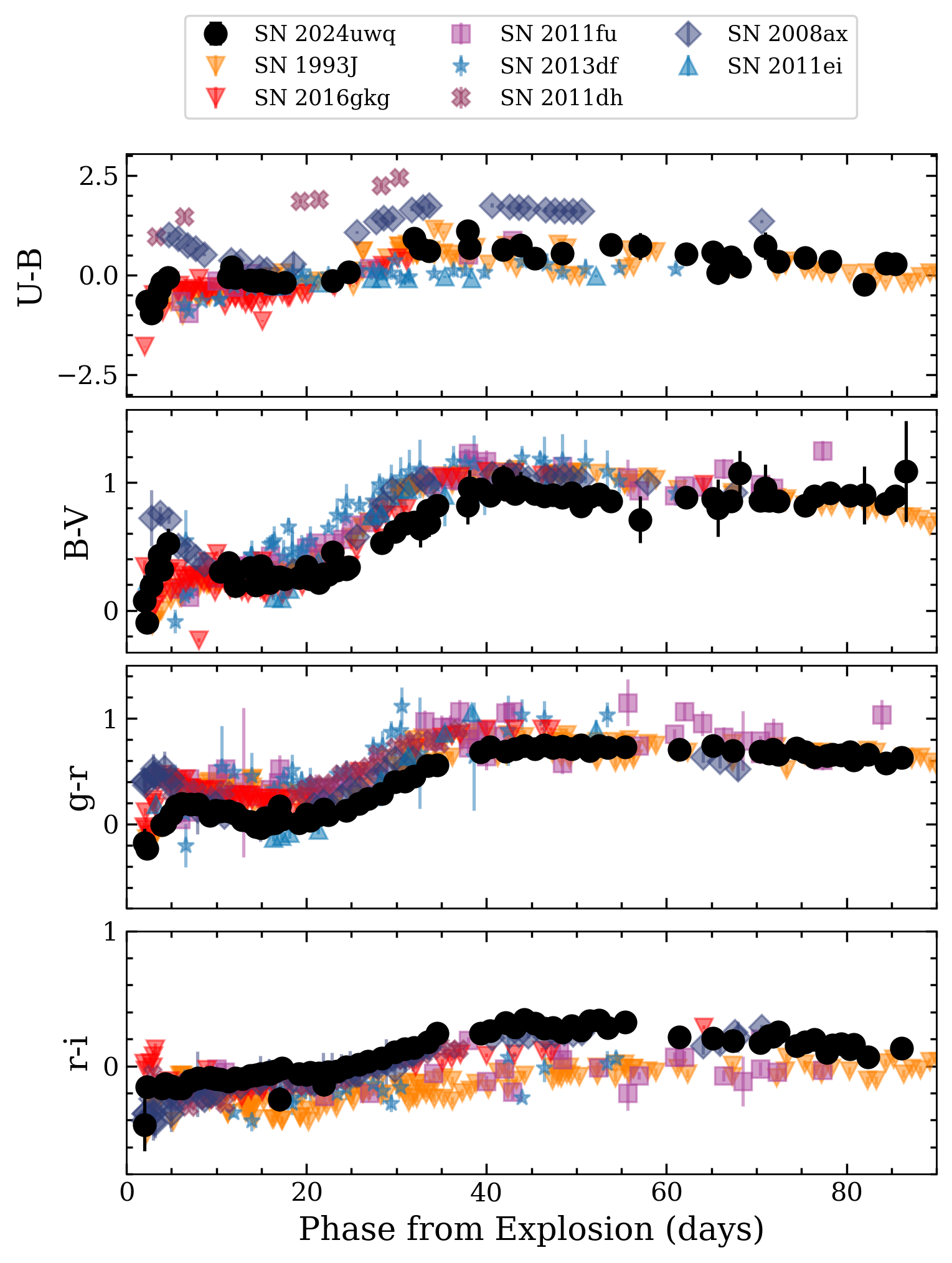}
    \caption{\textit{Left:} Absolute \textit{B}-band light curve of SN~2024uwq in comparison with Type IIb in the literature, with a zoom in on the earliest phases shown in the inset. \textit{Right:} Extinction corrected \textit{U} $-$ \textit{B}, \textit{B} $-$  \textit{V},  \textit{g} $-$  \textit{r}  and  \textit{r} $-$   \textit{i} color evolution of SN~2024uwq in comparison to the color evolution of typical Type IIb. We use the relationships prescribed in \citet{Jordi2006} for converting \textit{V} $-$ \textit{R}, \textit{R} $-$ \textit{I} to \textit{g} $-$ \textit{r} and \textit{r} $-$ \textit{i} respectively (when necessary). Data used in this figure are from \citealt{R1994} (SN~1993J), \citet{Pastorello2008} (SN~2008ax), \citet{A2011} (SN~2011dh), \citet{Morales2015_2011fu} (SN~2011fu), \citet{morales2014_2013df} (SN~2013df), and  \citet{Tartagalia2017}, \citet{Kilpatrick2022} (SN~2016gkg). }
    \label{fig:compare_colors}
\end{figure*}

\subsection{Spectroscopy}

A series of early high-cadence spectroscopic observations of SN~2024uwq was carried out using multiple facilities. Low-resolution optical spectra were acquired with the FLOYDS spectrograph mounted on the 2.0-m Faulkes Telescope South (FTS) at Siding Spring Observatory, Australia, through the Las Cumbres Observatory as part of the Global Supernova Project collaboration \citep{Brown_2013}. Observations were performed with a 2\arcsec\ wide slit aligned at the parallactic angle. One-dimensional spectra were extracted, reduced and calibrated according to standard procedures using the FLOYDS reduction pipeline \citep{Valenti_2014}.

Spectroscopic observations of SN~2024uwq were also acquired using the Robert Stobie Spectrograph (RSS) on the Southern African Large Telescope (SALT; \citealt{Smith_2006_SALT}). Data were reduced with a custom pipeline built in the PySALT package \citep{Crawford_2010}, incorporating standard processing steps such as bias subtraction, flat fielding, wavelength calibration using arc lamp exposures, and flux calibration with standard spectrophotometric stars. Additional optical spectra were also obtained with the Goodman High-Throughput Spectrograph (HTS) on the 4.1-m Southern Astrophysical Research Telescope (SOAR) for three epochs. Data reduction was performed using the Goodman HTS\footnote{\url{https://soardocs.readthedocs.io/projects/goodman-pipeline/en/latest/}} pipeline, employing standard reduction procedures.

NIR spectroscopy of SN~2024uwq was obtained using the Flamingos-2 instrument mounted on the Gemini-South telescope \citep{E2004, E2012}, as part of program GS-2024B-Q-215. The NIR observations were conducted on 2024 November 18 for the HK spectra (with an exposure time of 18 $\times$ 120,s at a relatively high airmass of 1.7) and on 2024 November 23 for the JH spectra (with an exposure time of 8 $\times$ 120,s). The data were reduced using custom IRAF scripts. Compared to GNIRS spectra from the Gemini-North telescope \citep[e.g.,][]{rho18sn}, Flamingos-2 spectra are less sensitive. 

\section{Photometry and Light Curve Evolution}\label{sec:lc_analysis}

\subsection{Light Curve with Early Shock Cooling Emission}

The multiwavelength light curve of SN~2024uwq, presented in Figure \ref{fig:lc_early}, reveals a distinct early-time emission excess followed by a rapid decline and a subsequent, more luminous second peak. The rise from first detection to the early excess is poorly constrained, with possible observations only in the ATLAS-\textit{o} band, while nearly all other bands fail to capture this phase, suggesting that the rise could be short-lived. The light curve then declines rapidly within the next $\sim$ 3 days, after which the optical magnitudes brighten again towards the second maximum. In the \textit{Swift} bands, after the initial decline, the rise to the second peak is less pronounced, with the \textit{UVM2} and \textit{UVW2} bands showing a flattening trend after the initial excess. We measure an absolute magnitude of M$_{B} = -16.3$ mag for the first observation in the \textit{B} band, which occurred approximately 2 days after the explosion. This is followed by a decline to M$_{B} = -15.7$ mag within $\sim$1.5 days of the initial peak. After +5 days, the light curve brightens again, reaching a more luminous secondary peak with an absolute magnitude of M$_{B} = -17.5$ mag around +20 days from the explosion epoch.

\begin{figure*}
    \centering
    \includegraphics[width=0.49\linewidth]{ 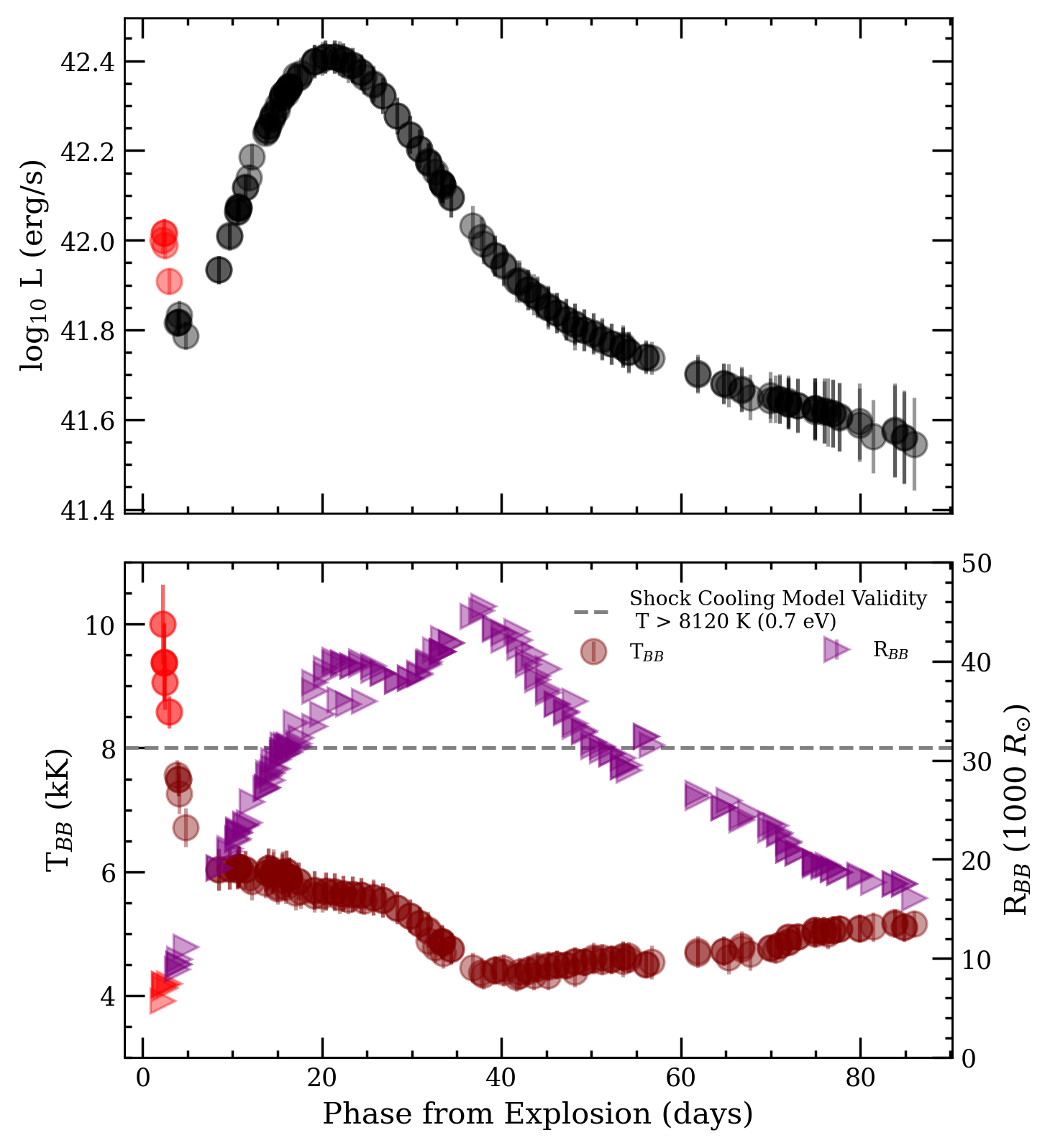}
    \includegraphics[width=0.49\linewidth]{ 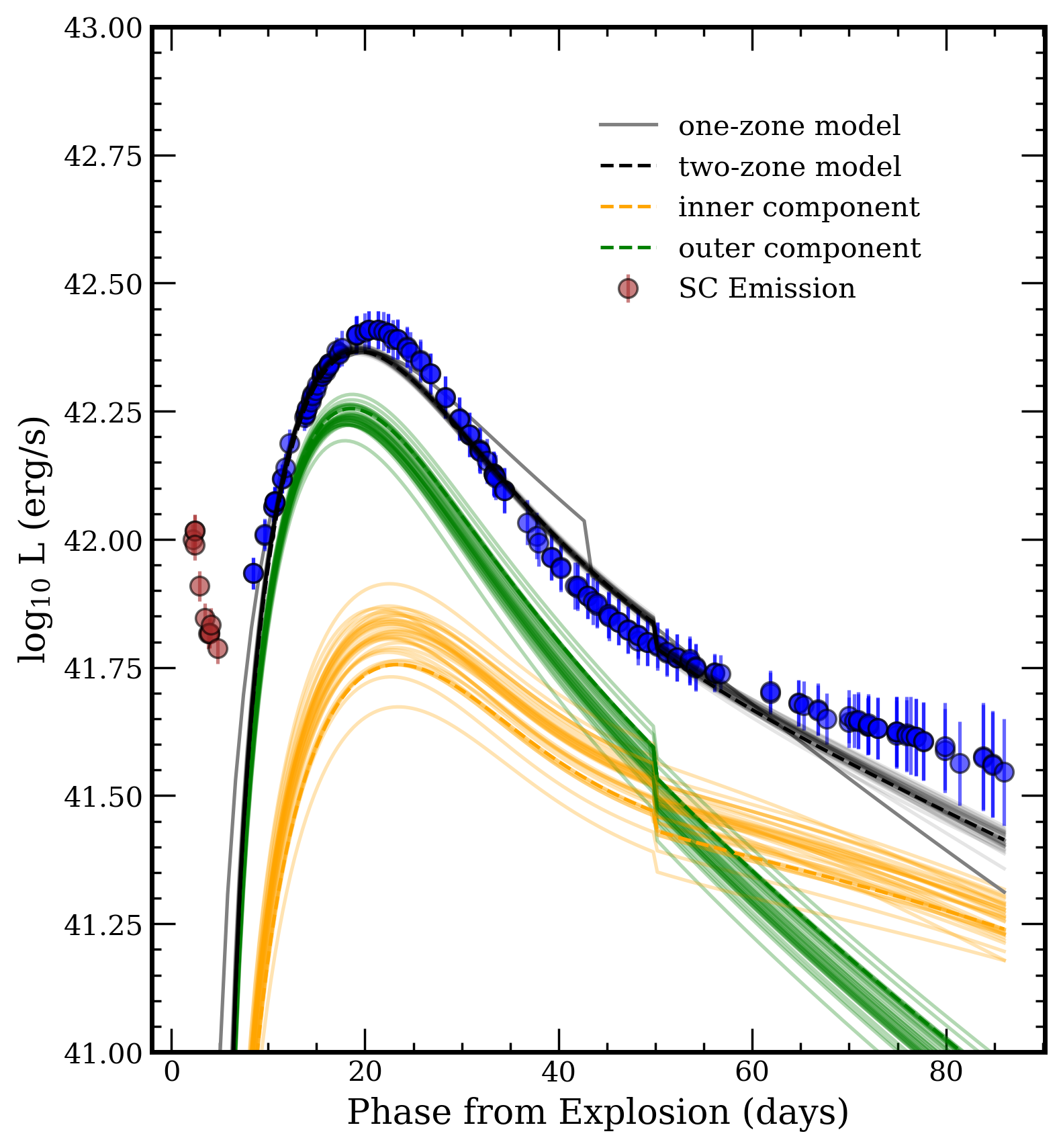}
    \caption{\textit{Left:} Bolometric light curve of SN~2024uwq along with the evolution of temperature and photospheric radius. We also mark in red, the earliest bolometric data whose multi-band observations are used for shock-cooling analysis. Shock Cooling models are valid only for data where the T$_{BB}$ is greater than 8120 K or 0.7 eV (see \citet{SW2017}) \textit{Right:} Bolometric light curve fit with a two-component Arnett model using Markov Chain Monte Carlo (MCMC). Black and grey lines depict two and one component models (150 random draws from the posterior); orange and green trace inner and outer ejecta contributions, respectively. Observational data are shown in blue with 1-$\sigma$ uncertainities.}
    \label{fig:bol_analysis}
\end{figure*}

We compare the absolute \textit{B}-band light curve of SN~2024uwq with other well studied Type IIb supernovae including SN~1993J \citep{R1994}, SN~2008ax \citep{Pastorello2008}, SN~2011ei \citep{Dan2013}, SN~2011dh \citep{A2011}, SN~2011fu \citep{Morales2015_2011fu}, SN~2013df \citep{morales2014_2013df} and SN~2016gkg \citep{Tartagalia2017}, as shown in Figure \ref{fig:compare_colors}. SN~2024uwq's early light curve shares close similarities with those of SN~1993J, SN~2011fu, SN~2013df, and SN~2016gkg, all of which exhibit characteristic early-time shock cooling emission. This early emission suggests an explosion originating from an extended progenitor star, contrasting with supernovae like SN~2008ax, SN~2011ei and SN~2011dh, which shows either a weak or absent early excess due to their more compact progenitors \citep{CS2010}. Although SN~2016gkg exhibits the most similar overall shape of the early light curve to SN~2024uwq in terms of decline and rise timescales, there are notable differences in the early excess and second peak luminosities. The initial emission excess observed in SN~2016gkg is significantly more luminous than SN~2024uwq at comparable early epochs \citep{Tartagalia2017, Kilpatrick2022}.
In contrast, as shown in Figure \ref{fig:compare_colors}, the second peak of SN~2024uwq reaches a higher luminosity than that of SN~2016gkg.
When comparing SN~2024uwq, it is crucial to acknowledge that despite ATLAS pre-detections, its true early excess peak remains uncertain due to observational cadence and potential band-dependent emissions at these earliest phases.

We measure the \textit{B}-band apparent magnitude decline rate for SN~2024uwq after the initial maximum to be $\sim$ 0.64 mag$/$days over the first 5 days. This decline is slower than that of SN~2016gkg, which showed a steeper decline of 0.81 mag/day over $\sim$ 2 days, but faster than SN~1993J's more gradual decline of 0.31 mag/day over 5 days (see Figure \ref{fig:compare_colors}). SN~2024uwq, like SN~2016gkg and SN~2011fu, exhibits a second maximum that is brighter than or comparable to the initial maximum, a characteristic that differentiates them from SN~1993J and SN~2013df, whose secondary peaks are significantly less luminous. Following the second maximum, SN~2024uwq demonstrates a relatively slow decline, comparable only to SN~2011fu, while most other SNe in the sample show much faster timescales of decline. 
The slow decline rate observed after the second maximum is consistent with radioactive heating from the decay of $^{56}\mathrm{Ni}$, which powers the later light curve phases in all Type IIb supernovae. For SN~2024uwq, the particularly gradual decline suggests a relatively high $^{56}\mathrm{Ni}$ or enhanced trapping of gamma rays, as discussed later in \ref{subsec: bol_analysis}.

\subsection{Color Evolution}

In Figure~\ref{fig:compare_colors}, we present the extinction-corrected color evolution of SN~2024uwq in \textit{U} $-$ \textit{B}, \textit{B} $-$ \textit{V}, \textit{g} $-$ \textit{r}, and \textit{r} $-$ \textit{i}, compared to well-studied Type IIb supernovae.  SN~2024uwq lacks the early red excess in \textit{U} $-$ \textit{B}, as observed for SN~2008ax and SN~2011dh, which are associated with compact progenitors dominated by radioactive heating even at earlier phases \citep{Pastorello2008, A2011}. The \textit{U} $-$ \textit{B} colors of SN~2024uwq closely resemble that of SN~1993J, SN~2013df and SN~2016gkg.  

The \textit{B} $-$ \textit{V} and \textit{g} $-$ \textit{r} colors of SN~2024uwq initially exhibit a blueward evolution until the second maximum ($\sim 20$ days) before transitioning to a reddening phase. This behavior is consistent with most SNe IIb, though SN~2024uwq maintains a systematically bluer color nearly up to 50 days past explosion. The \textit{r} $-$ \textit{i} evolution of SN~2024uwq shows gradual reddening up to 60 days, followed by a slight blueward trend, with values higher than those of SN~1993J and SN~2011fu but comparable to SN~2016gkg and SN~2008ax \citep{R1994, Pastorello2008, Morales2015_2011fu, Tartagalia2017}, see Figure \ref{fig:compare_colors}.

\subsection{Bolometric Analysis}\label{subsec: bol_analysis}

We used photometric measurements corrected for extinction in all available passbands to construct the bolometric light curve of SN~2024uwq, employing \texttt{SuperBol}\footnote{\url{https://github.com/mnicholl/superbol}} \citep{Nicoll2018}. Given the critical role of UV observations in constraining blackbody fits, but its sparser sampling during the rapidly evolving shock-cooling and subsequent rise, we interpolated light curves using higher-order polynomials, approximately fifth to eighth order, with the specific order varying by band based on light curve evolution. Each observed epoch was fit to a blackbody spectral energy distribution, facilitating the calculation of bolometric luminosities, blackbody temperatures, and blackbody radii as functions of time. The bolometric light curve of SN~2024uwq along with temperature and photospheric radius are shown in Figure \ref{fig:bol_analysis}. We also mark the shock-cooling measurements with red that are valid to be used for early emission modeling as described in Section \ref{sec:sc_model}. These measurements last $\sim 3$ days, where shock cooling emission dominates the luminosity over radioactive decay. We make a validity cut on the time range based on the temperatures $\geq 8120$ K (0.7 eV), as described in \citet{SW2017} and \citet{MSW2023} (see Equation A3).

The bolometric luminosity of the first observation of SN~2024uwq is measured to be $\log L \, (\mathrm{erg\,s^{-1}}) = 42.017 \pm 0.031$. Following the initial shock cooling phase, the bolometric luminosity reaches a second maximum, measured at $\log L \, (\mathrm{erg\,s^{-1}}) = 42.409 \pm 0.037$.
The effective temperature of SN~2024uwq evolves from a temperature of \textit{T} $\sim$ 10 kK during the first observation, with a fast decrease to 6 kK up to $\sim 10$ days and reaching around 5.6 kK during the second light curve maximum. The temperature evolution is similar to that of SN~2016gkg where the initial observation yielded \textit{T} $\sim$ 13 kK, rapidly decreasing to 7.9 kK \citep{Tartagalia2017}. The photospheric radius of SN~2024uwq during the initial peak is estimated to be $R_\mathrm{phot} \sim 7000 R_\odot $, expanding to 
$ \geq 4 \times 10^{4} R_\odot $ at approximately 40 days after which it gradually recedes as SN~2024uwq further evolves.

\subsection{$^{56}$Ni Mass Estimates}

We model the second peak of the bolometric light curve of SN~2024uwq using both one and two component variations of the  analytical framework adapted from \citet{Arnett_1982} and \citet{Valenti_2008}, originally developed for Type Ia supernovae (SNe Ia) and subsequently extended to SESNe \citep{Lyman_2016,Dong2024}. The one-component model assumes a single homogeneous ejecta structure, where the photospheric phase luminosity \( L_{\text{bol,phot}}(t) \) is powered by radioactive decay of \( ^{56}\text{Ni} \) and \( ^{56}\text{Co} \), with gamma-ray leakage \(\Gamma(z)\) integrated over time (Equation 4; \citet{Valenti_2008}). The characteristic time scale, \( \tau_m \propto \left(M_{\text{ej}}^3 / E_{\text{k}}\right)^{1/4} \) depends on the mass of the ejecta \( M_{\text{ej}} \), the kinetic energy \( E_{\text{k}} \), and the constants \( \kappa_{\text{opt}} = 0.07 \, \mathrm{cm}^2 \, \mathrm{g}^{-1} \), \( \beta = 13.8 \).  The two-component model for SESNe is motivated by the inability of single-zone models to reconcile photospheric phase luminosity (dominated by outer low-density ejecta with rapid cooling) and nebular phase emission (powered by inner dense ejecta with enhanced gamma ray trapping), as well as the need for $^{56}$Ni mixing observed in SESNe. Here, \( L_{\text{bol,tot}}(t) \) becomes the sum of contributions from both components:  
\begin{equation}
L_{\text{bol,tot}}^{\text{2-comp}}(t) = L_{\text{phot}}(t) + L_{\text{neb}}(t),
\end{equation}  
where \( L_{\text{phot}}(t) \) and \( L_{\text{neb}}(t) \) retains the formalism described in \citet{Valenti_2008} and \citet{Chat2012}.

\begin{figure*}
    \centering
    \includegraphics[width=\textwidth]{ 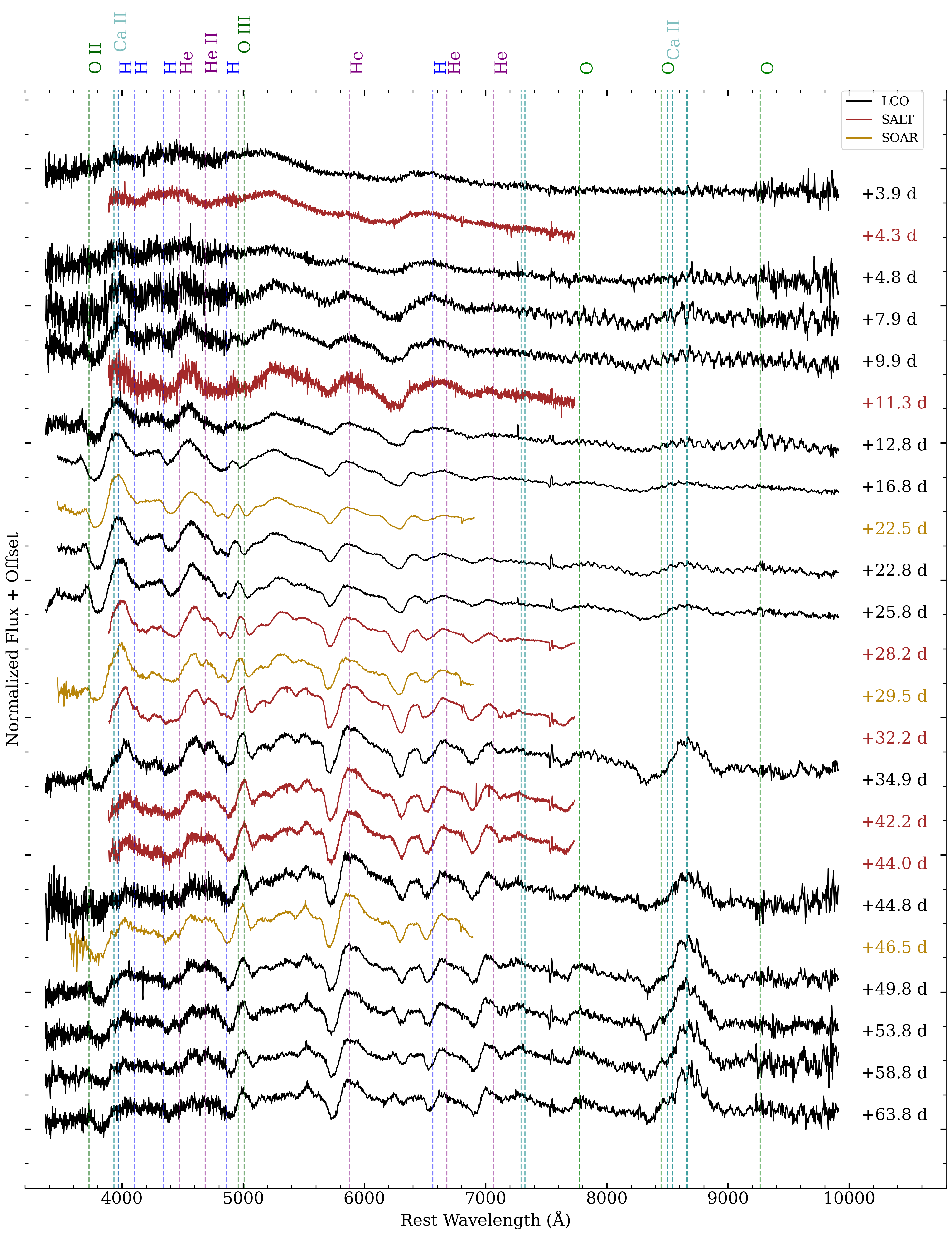}
    \caption{Optical Spectra of SN~2024uwq showing temporal evolution from +4 days to +64 days, with respect to our assumed explosion epoch of MJD 60558.63. The most prominent lines in the spectra are identified.}
    \label{fig:spec}
\end{figure*}

We constrain the time when shock cooling ends with respect to explosion epoch $t_s$, ejecta mass $M_{ej}$, nickel mass $M_{56Ni}$, kinetic energy $E_k$, and inner mass fraction $f_{inner}$ using MCMC sampling adopting uniform priors for the physical parameters. We fit the bolometric light curve from $-10$ to $+15$ days and $>60$ days after maximum, epochs that best represent photospheric and nebular phases of SN~2024uwq's evolution. $t_s$ is anchored to the first data point used in the fit - which includes only data after the shock cooling phase, yielding shock cooling end time to be $t_s = 4.94^{+0.042}_{-0.079}$ days, consistent with our assumed explosion epoch (see Figure \ref{fig:bol_analysis}). The best-fit parameters give $M_{56Ni} = 0.098^{+0.001}_{-0.001} \, M_\odot$, $M_{ej} = 3.00^{+0.103}_{-0.089} \, M_\odot$, $E_{k, \text{total}} = 2.75^{+0.02}_{-0.03} \times 10^{51}$ erg, and $f_{inner} = 0.33 \pm 0.03$, consistent with SESNe population studies \citep{Lyman_2016, Taddia_2018}. Simplified one- and two-zone models struggle to match the observed luminosity beyond 60 days (Figure~\ref{fig:bol_analysis}). While increasing the inner ejecta density component might explain the late-time emission, it would also broaden the primary peak, highlighting the limitations of these simplified models compared to a continuous density profile. Alternative energy sources, such as circumstellar medium interaction \citep[e.g.,][]{Chevalier1994, Moriya2023, Rizzo2023} or magnetar spin-down \citep[e.g.,][]{Kasen2010, Woosley2010}, have been suggested to account for the extended luminosity tail in CCSNe.

\section{Spectroscopy}\label{sec:spectra}

\subsection{Optical Spectra}

We show the spectral evolution of SN~2024uwq
ranging from +4 d to +64 d after explosion in Figure \ref{fig:spec}. The most prominent lines are identified in the figure. In the earliest spectrum, taken at +4 days, a broad P-Cygni H$\alpha$ profile is evident, along with an absorption feature of H$\beta$ near 4700 \AA. Both the H$\alpha$ and H$\beta$ absorption components exhibit flat-topped profiles, indicative of an expanding hydrogen shell. These early features are also seen in the spectra of SN~1993J, SN~2013df and SN~2016gkg at similar phases \citep{morales2014_2013df, Tartagalia2017}. Additionally, the spectrum shows He I 5876 \AA\ absorption, as well as weak traces of Ca II H $\&$ K absorption at 3934 \AA\ and 3968 \AA. The low continuum temperature at this epoch, derived as 8500 K from a blackbody fit (see Fig. \ref{fig:bol_analysis}), combined with the presence of low ionization elements, suggests rapid cooling following shock breakout \citep{A2011}.

\begin{figure}
\includegraphics[width=\linewidth]{ 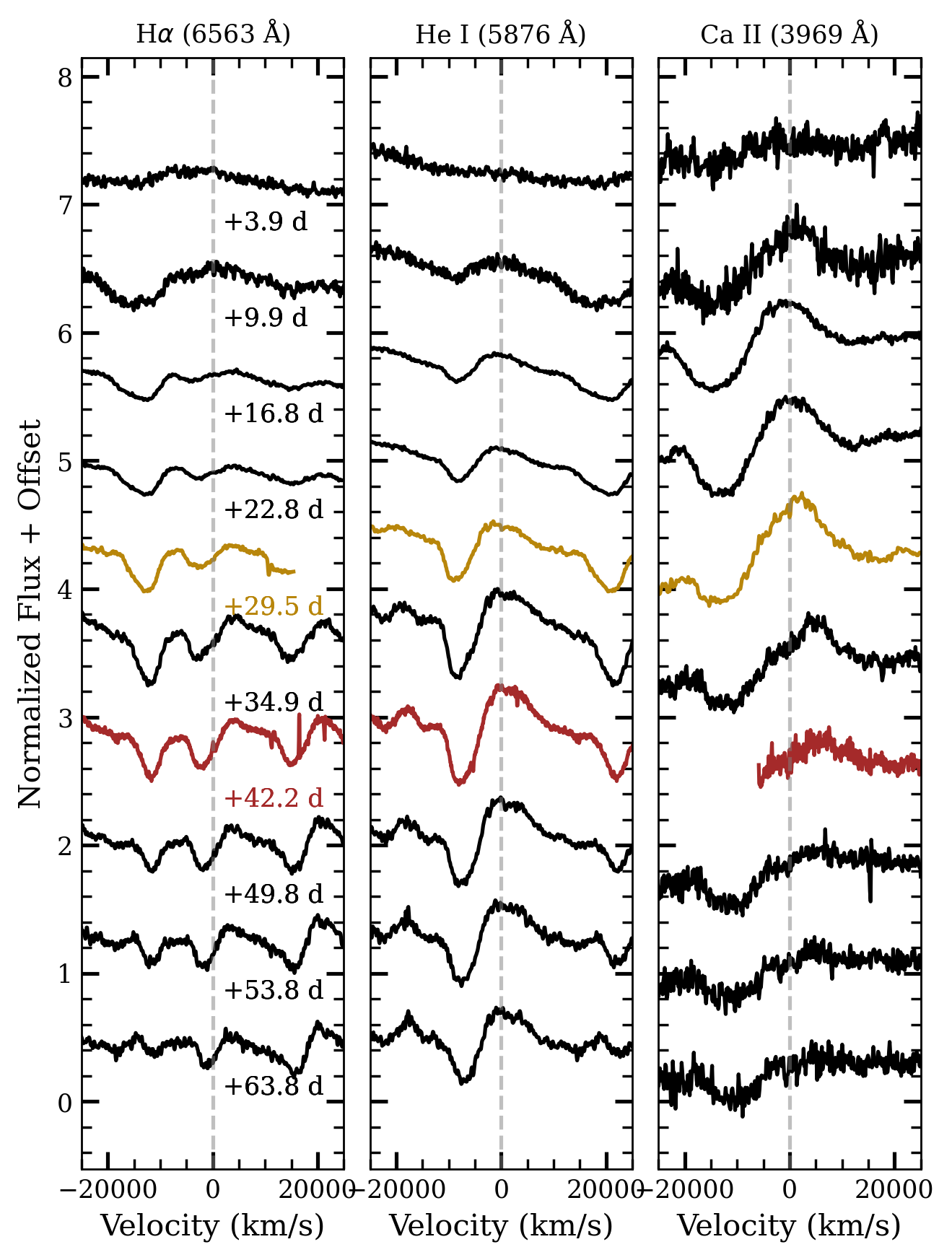}
\includegraphics[width=\linewidth]{ 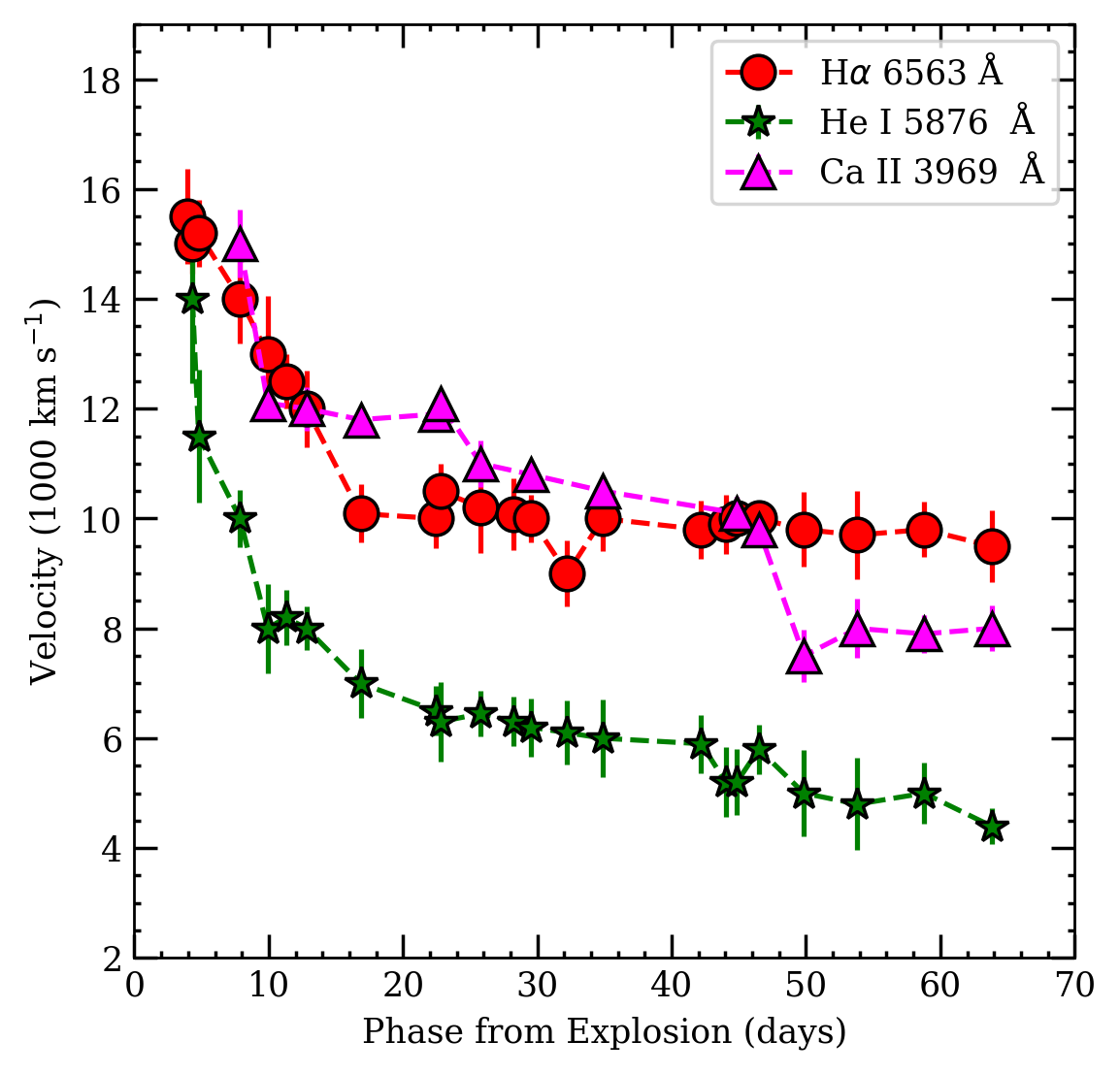}
\caption{\textit{Top:} Multi-epoch spectral evolution of H$\alpha$ (6563~\AA), He~\textsc{i} (5876~\AA), and the Ca~\textsc{ii} H \& K doublet (3969~\AA), observed between $+4$ to $+64$ days. The H$\alpha$ emission line profile evolution is strongly affected by He I 6678 line emergence \textit{Bottom:} Velocity evolution derived from P-Cygni minima of H$\alpha$ (red), He~\textsc{i} (green), and Ca~\textsc{ii} H \& K (magenta), plotted relative to explosion epoch.}
   \label{fig:velocity_evolution}
\end{figure}

The evolution of the H$\alpha$, He I, and Ca II H \& K lines across multiple epochs is presented in Figure \ref{fig:velocity_evolution}. The H$\alpha$ line strengthens as SN~2024uwq evolves, and by +15 days, a secondary component, likely He I 6678 \AA, emerges, similar to the evolution of SN~2016gkg \citep{Tartagalia2017}. Strong P-Cygni profiles of He I lines develop following the shock-cooling phase and intensify by +23 days. The Ca II H \& K P-Cygni features are first discernible around +10 days, exhibiting steady growth in subsequent epochs, while the Ca II NIR triplet shows a pronounced increase in strength starting at +34 days.

At epochs greater than +40 days, the H$\alpha$ line intensity decreases by approximately a factor of two, whereas the He I lines, including 5876 \AA, 6678 \AA, and 7065 \AA, strengthen significantly, with an enhancement factor of $\sim$2 relative to earlier spectra. The He I 6678 \AA\ and 7065 \AA\ features, absent in the early spectra, become distinctly visible at this stage, consistent with previous observations of SN~2011fu, SN~2013df and SN~2016gkg (\cite{Morales2015_2011fu, morales2014_2013df, Tartagalia2017}.

\subsection{NIR Spectroscopy}

\begin{figure}
   \centering
\includegraphics[width=\linewidth]{ 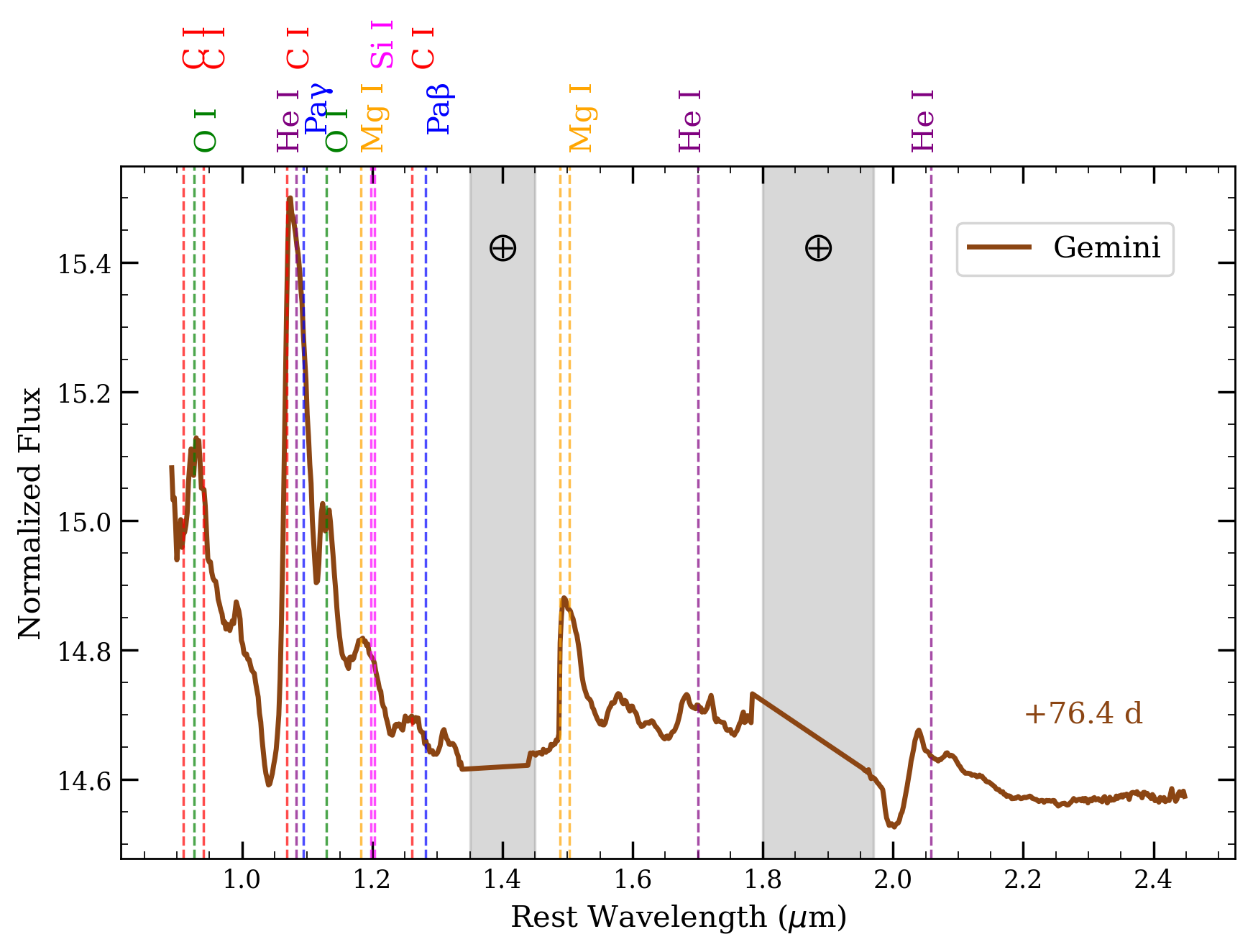}
\includegraphics[width=\linewidth]{ 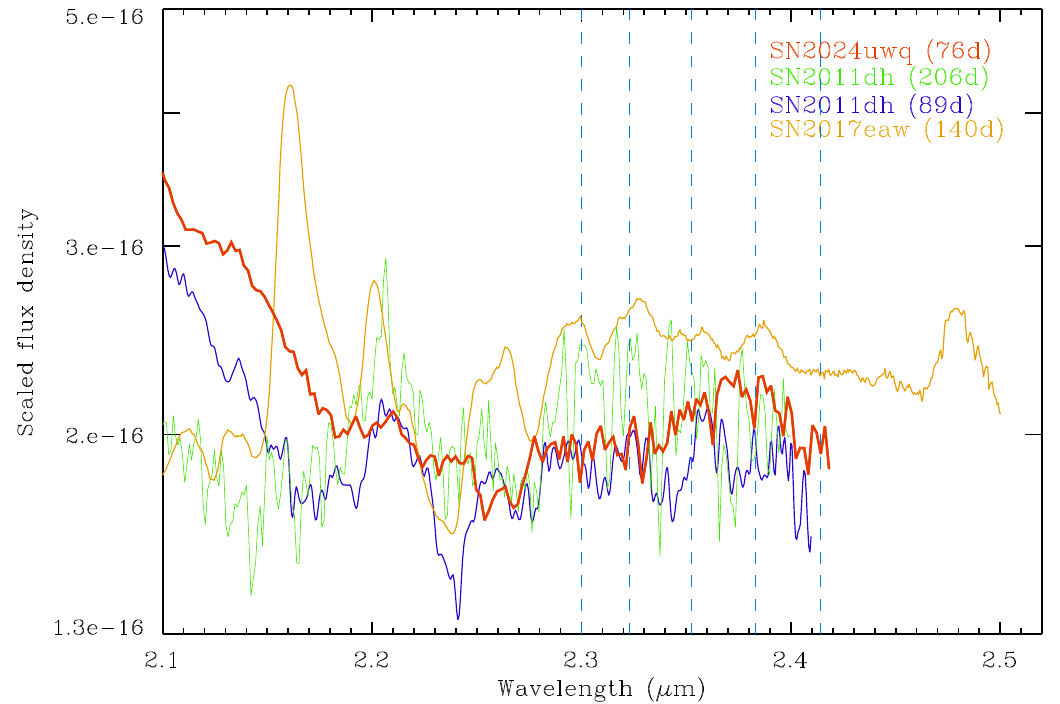}
    \caption{ (Top: a) NIR spectra of SN~2024uwq at +76 days obtained using Gemini F-2. The gray bands mark the regions with high telluric absorptions and emissions. (Bottom: b) The CO first overtone feature of SN~2024uwq (in red) shows strong resemblance to those observed in the Type IIb SN~ 2011dh and is also compared with those of the Type IIP SN~ 2017eaw. The vertical dashed lines mark the bandheads of  $\Delta v = 2$ transitions.}
   \label{fig:nir_spectra}
\end{figure}

We present a NIR spectrum of SN\,2024uwq at +76 days in Figure \ref{fig:nir_spectra}. The strongest feature in the spectrum is a P-Cygni profile peaking around $\lambda$1.0830 $\mu$m, which is due to He~I. The concurrent presence of a clear P-Cygni profile around $\lambda$2.0581 $\mu$m confirms a significant contribution from He~I, which is expected in an evolved spectrum of a SN IIb \citep[e.g.,][]{Taubenberger2011, Bufano2014, Ergon2015, Shahbandeh2022}. 
The He~I P-Cygni profile at  $\lambda$1.0830 $\mu$m is potentially contaminated by  Pa$\gamma$ ($\lambda$1.094 $\mu$m), while the Pa$\beta$ ($\lambda$1.282 $\mu$m) absorption is not evident. SN\,2024uwq exhibits absorption features at $\lambda 0.9264$ and $\lambda 1.129\,\mu$m, consistent with O\,\textsc{i} lines commonly seen in stripped-envelope supernovae \citep{Shahbandeh2022}. Several C\,\textsc{i} lines are also identified in the NIR spectra of SN\,2024uwq. While weaker C\,\textsc{i} features at $\lambda 0.9093\,\mu$m, $\lambda 0.9406\,\mu$m may be blended with the nearby O\,\textsc{i} $\lambda 0.9264\,\mu$m line, the prominent C\,\textsc{i} $\lambda 1.0693\,\mu$m feature is detected and likely contributes significantly to the broad P-Cygni along with He\,\textsc{i} $\lambda 1.0830\,\mu$m. Most evolved NIR spectra of stripped-envelope SNe show an emission-like Mg~I feature around 1.5 $\mu$m \citep[with contributions from Mg~I $\lambda$1.4878 $\mu$m and Mg~I $\lambda$1.5033 $\mu$m;][]{Shahbandeh2022}. This feature is observed in SN\,2024uwq. The emission band at $\sim$1.19 $\mu$m can likely be attributed to Si I $\lambda \lambda$1.198, 1.203 $\mu$m blended with Mg~I $\lambda$1.183 $\mu$m as observed for SN~IIb~2011hs \citep{Bufano2014}. Similarities with other SNe IIb at NIR wavelengths further confirms the classification of SN\,2024uwq.

Figure~\ref{fig:nir_spectra} shows that the first overtone CO (2.25-2.45 $\mu$m) is likely detected in the NIR spectra of SN~2024uwq. The presence of the first overtone CO feature indicates the formation of CO molecules as the ejecta cool, a process that can subsequently lead to dust formation. The CO feature in SN~2024uwq bears some resemblance to that observed in the Type IIb SN~2011dh \citep{Ergon2015}. In SN~2011dh, the CO emission observed at 206 days post-explosion was inferred to have a temperature of approximately 2300K and an expansion velocity of 1500~\kms\ \citep{Ergon2015}. Compared to the Type IIP SN~2017eaw, the CO bandheads in SN~2024uwq appear less distinct.

\subsection{Velocity Evolution}
Figure \ref{fig:velocity_evolution} shows the evolution of the expansion velocities of H$\alpha$, He I, Ca II H $\&$ K for SN~2024uwq.
The expansion velocities are derived from the position of the minima of the P-Cygni profile of each respective line.
The H$\alpha$ velocity starts as high as $\sim$ 15,500 km s$^{-1}$ at + 4 days, subsequently decreasing to $\sim$ 10,000 km s$^{-1}$ at +30 days. For our earliest phases, we measure the minimum of the P-Cygni of the H$\alpha$ line by fitting a Gaussian to the absorption profile, and derive H$\alpha$ expansion velocities, which is close to the values obtained for other IIb SNe at similar phase including SN~2008ax, SN~2013df, SN~2011dh and SN~2016gkg \citep{Pastorello2008, morales2014_2013df, A2011}.
For instance, SN~2016gkg H$\alpha$ line profiles evolve in a similar range with expansion velocities declining from ~16,500 km s$^{-1}$ at +1.70 days to ~12,200 km s$^{-1}$ at +21 days \citep{Tartagalia2017}. The He I 5876 \AA\ expansion velocities evolve much faster than the H$\alpha$. First appearing around +4 days, the expansion velocity decreases steeply from 14,000 km s$^{-1}$ to 7000 km s$^{-1}$ by +15 days. With an initial rapid drop at earlier phases, the Ca II H \& K expansion velocities then evolve steadily, tracking the H$\alpha$ velocity evolution until +46 days, after which they decrease below  H$\alpha$. For later epochs, the rate of velocity change for all three lines decreases. The H$\alpha$ expansion velocities of SN~2024uwq are systematically higher than those of normal Type II SNe (e.g., $\sim$8,000-12,000 km s$^{-1}$ at comparable epochs; \citealt{MS2024_23axu, Andrews_2024}), consistent with the enhanced ejecta velocities observed in SESNe.

\subsection{Comparison to other SN IIb}

\begin{figure}
   \centering
\includegraphics[width=\linewidth]{ 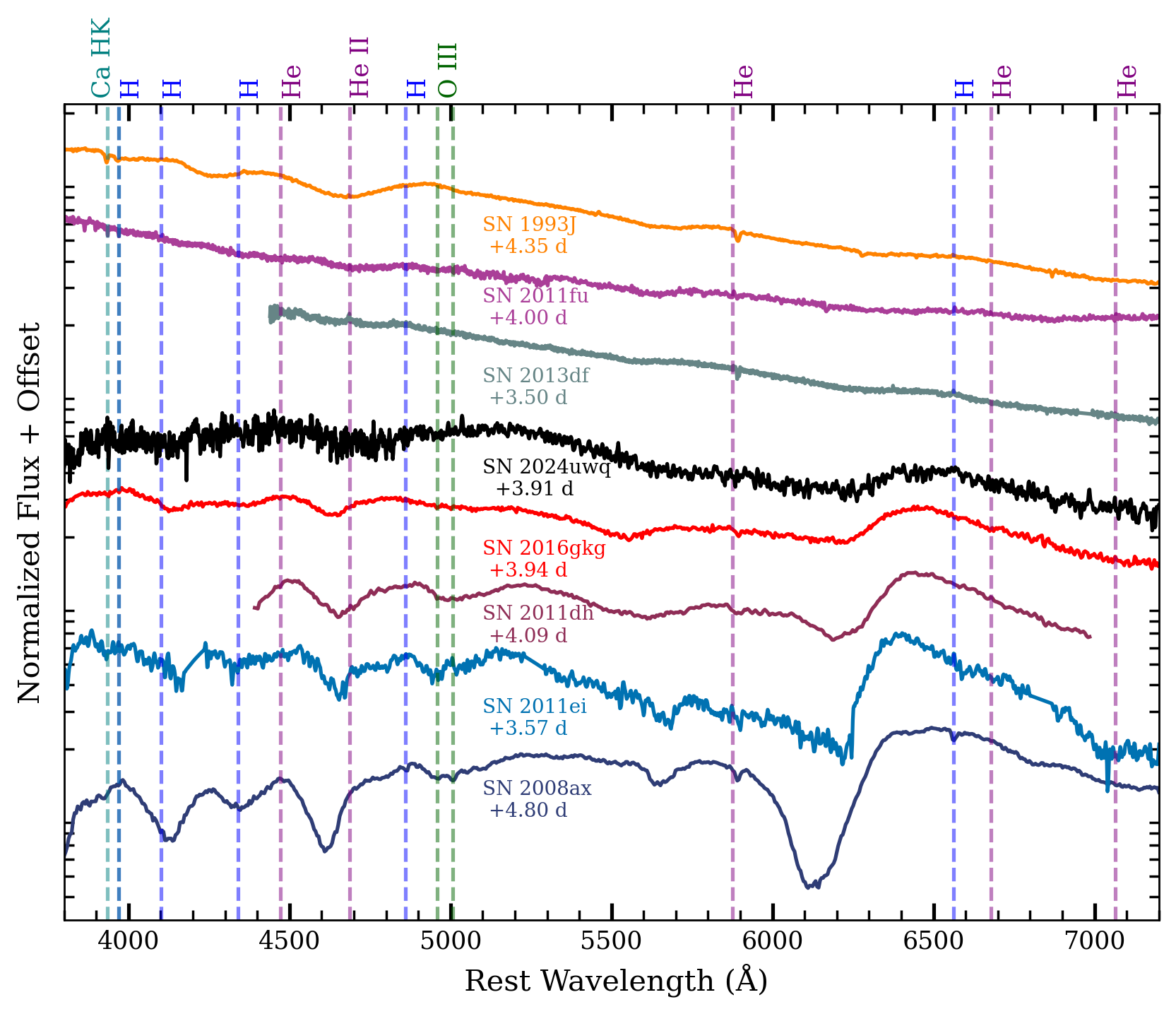}
\includegraphics[width=\linewidth]{ 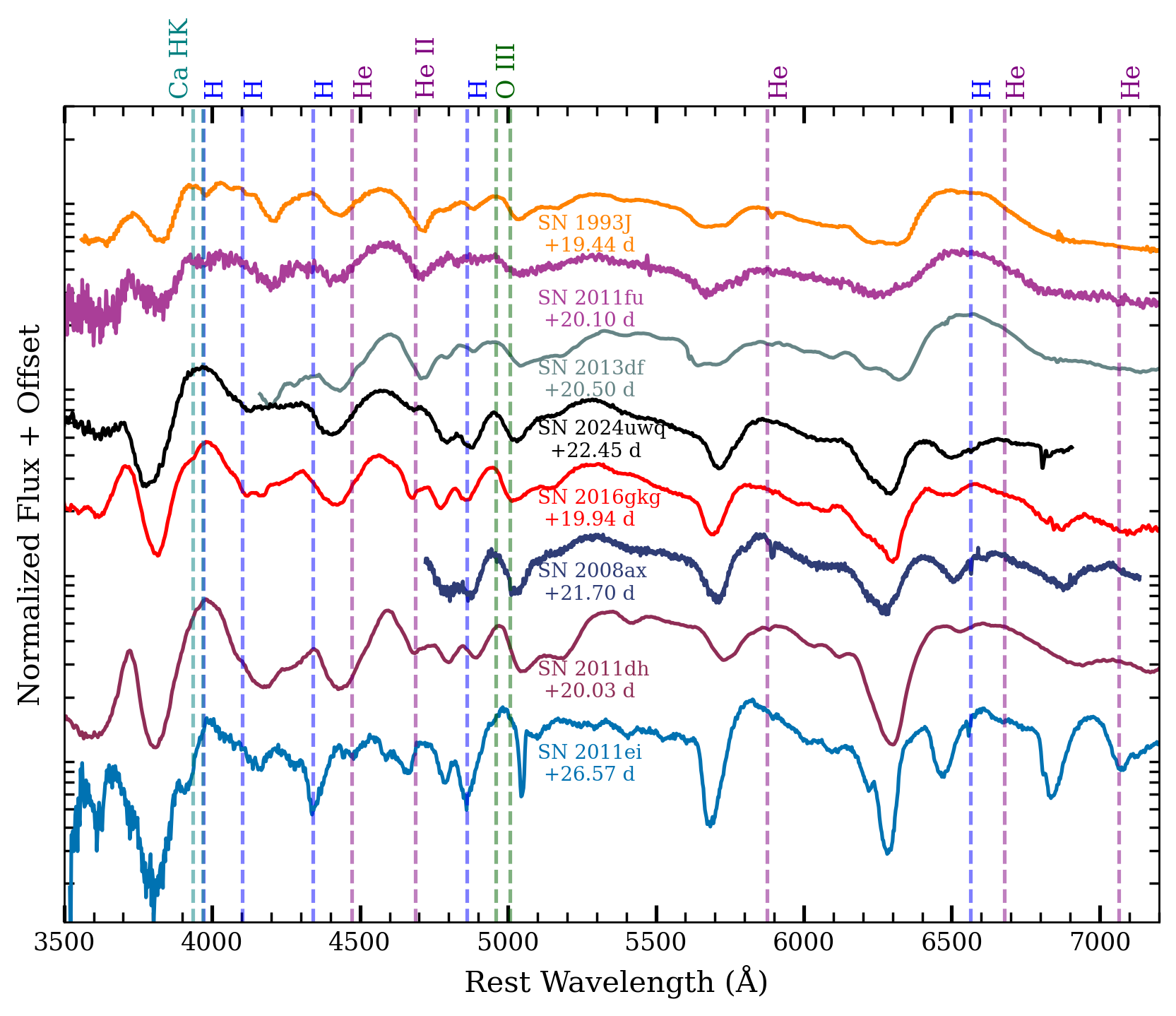}
\includegraphics[width=\linewidth]{ 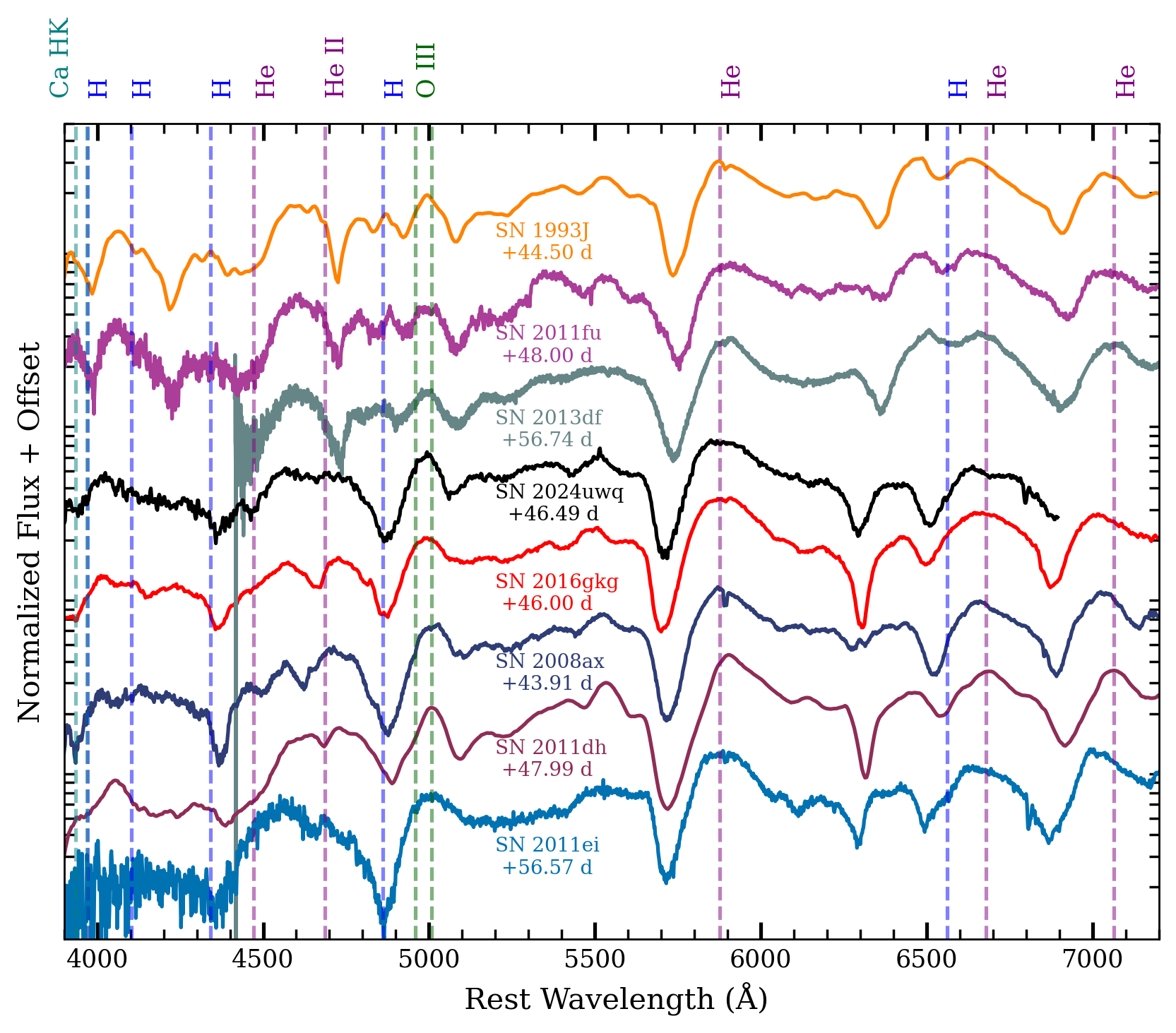}
    \caption{Comparison of the optical spectra of SN~2024uwq at +4 days (\textit{top}), +20 days (\textit{middle}) and +46 days (\textit{bottom}) with other Type IIb. The spectra have been redshift corrected and shifted vertically for clarity.}
   \label{fig:compare_spectra}
\end{figure}

We compare the optical spectra of SN~2024uwq at +4, +20 and +46 days with that of SN~1993J, SN~2008ax, SN~2011fu, SN~2011ei, SN~2011dh, SN~2013df and SN~2016gkg at similar phases in Figure \ref{fig:compare_spectra}. All data for the Type IIb SNe comparison plot were downloaded from WISeRep \citep{Yaron2012}. The early-phase spectra of Type IIb supernovae, shown in the top panel, exhibit significant diversity. SNe~1993J, 2011fu, 2013df display a blue, almost featureless continuum with shallow hydrogen and helium lines \citep{Mathesan2000, Morales2015_2011fu, morales2014_2013df}. In contrast, SN~2008ax, SN~2011ei and SN~2011dh are redder, characterized by stronger ``saw-toothed" H$\alpha$ spectral features, which indicate the absence of a shock-cooling emission phase and possibly a compact progenitor \citep{Pastorello2008, Dan2013, A2011}. SN~2024uwq evolves relatively similar to SN~2016gkg, showing  a blue-continuum and stronger Balmer features in its earliest spectra compared to SNe~1993J, 2011fu, and 2013df \citep{Tartagalia2017}.

At +20 days, all spectra exhibit more prominent He I 5876~\AA\ and Balmer features, as shown in the middle panel. This phase is close to the secondary peak for SN~2024uwq, after the initial decline in the light curve due to shock cooling.  SN~2024uwq and SN~2016gkg display stronger He I features compared to other SNe, except for SN~2011ei and SN~2008ax, that show the strongest He I P-Cygni profiles within just $\sim$ weeks of explosion \citep{Chornock2011,Dan2013, Tartagalia2017}. The line profiles of SN~2011fu evolve coherently, resembling those observed in SN~1993J and SN~2013df at this phase. The H$\alpha$ line profile in SN~2024uwq and SN~2016gkg at this phase starts to develop two components, contrasting with the single component profiles observed in SN~1993J, SN~2011fu, and SN~2013df, possibly due to the development of the strong nearby He I 6678 \AA. At this phase, blueshifted H$\alpha$ absorption is strongest in SN~2011dh, but the secondary absorption component is less pronounced compared to SN~2008ax, SN~2016gkg, and SN~2024uwq. Strong Ca II H $\&$ K
features develop for SN~2024uwq, SN~2016gkg and SN~1993J, however the absorptions in SN~2011ei and SN~2011dh remains the strongest among the sample \citep{Barbon1995,Dan2013, Ergon2014, S2019}. 

The bottom panel shows the spectra of all SNe at around +46 days, at which point they start to become redder. He I is the dominant line profile in the spectra with weakening strengths of H$\alpha$. SN~2008ax shows a weak or no blue-shifted H$\alpha$ component at this phase. SN~2011fu shows the least strength in He I and H$\alpha$ as compared to the other SNe in the sample. The O [III] 4959, 5007 \AA\ lines in all SNe strengthen compared to +20 days. SN~2016gkg shows broader He I profiles as compared to SN~2024uwq indicating higher expansion velocities.  All SNe also show He I 6678 \AA\ and 7065 \AA\ in their optical spectra at this phase \citep{Mathesan2000, Modjaz2014,  S2019}. Overall, SN~2024uwq's spectral evolution closely resembles that of SN~2016gkg and SN~2013df, both of which exhibit early shock-cooling signatures, in contrast to SNe~2008ax, SN~2011ei, and SN~2011dh, which lack this early light curve feature.
\section{Shock Cooling Emission Modeling}\label{sec:sc_model}

\begin{figure*}
    \centering
    \includegraphics[width=\textwidth]{ 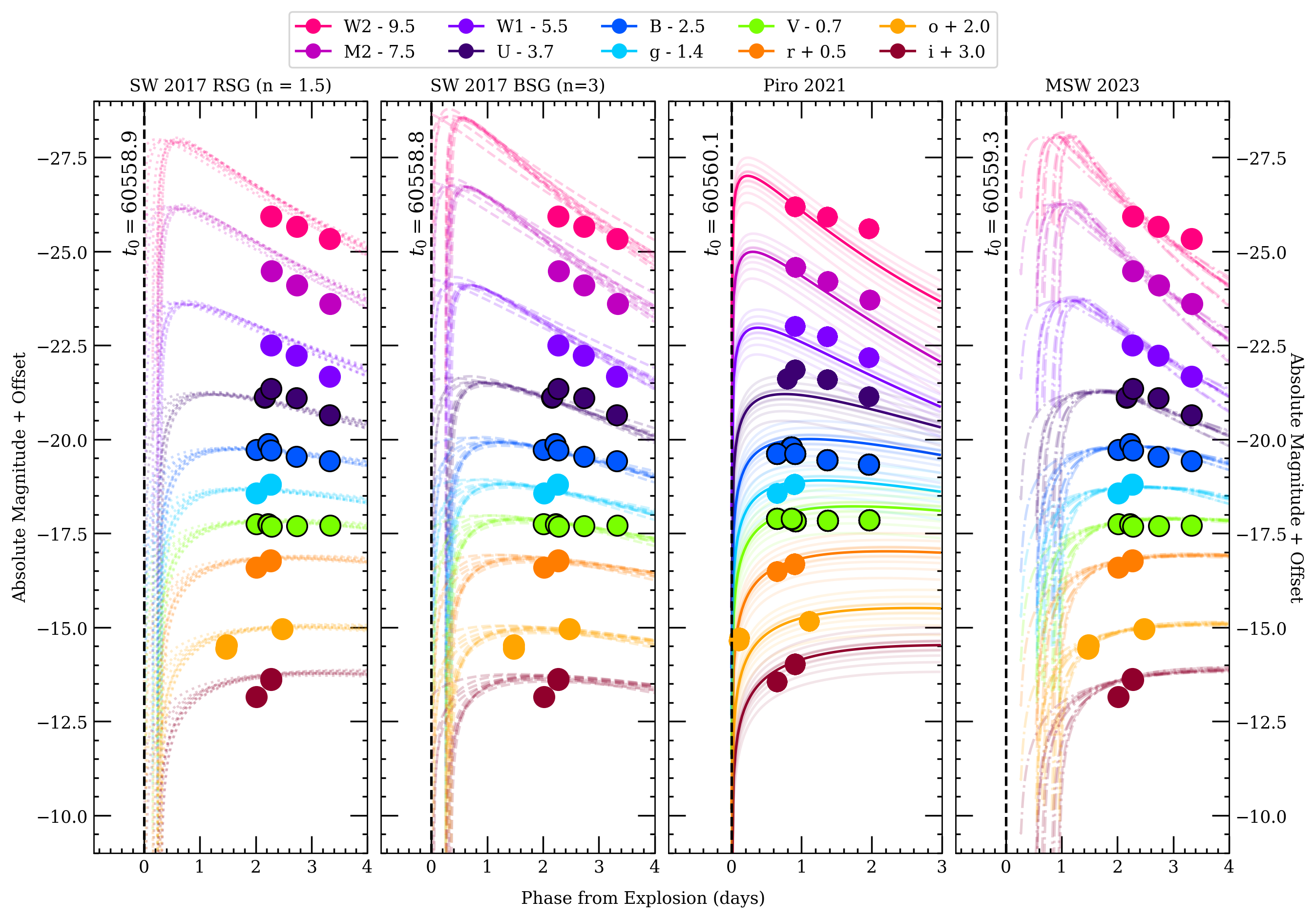}
    \vspace{-2em}
    \caption{Early-time light curve of SN 2024uwq with shock cooling model fits. Each panel represents the individual shock cooling model. The family of model light curves in each panel represents 50 models randomly sampled from the derived posterior probability distribution in individual bands. The best-fit values and prior distributions of the physical parameters for each model considered in this work are listed in Table \ref{tab:parameters}.}
    \label{fig:sc_early}
\end{figure*}

Several models have been developed over a decade to analytically describe the early shock-cooling emission in core-collapse supernovae, encompassing both the ``planar" and ``spherical" phases of the cooling. The planar phase, where the emitting shell is significantly thinner than the stellar radius, has been modeled through exact and approximate analytic solutions \citep{SKW2011, Katz2012, SNK2013}. As the emission progresses into the spherical phase, characterized by shell expansion to radii much larger than the star, solutions have been provided by \citet{Nakar2010}, \citet{Rabinak2011}, \citet{Nakar2014}, \citet{Piro2015}, \citet{SW2017}, \citet{Shussman2016}, and \citet{Piro2021}.

\begin{table*}[ht]
\centering
\setlength{\tabcolsep}{1pt} 
\renewcommand{\arraystretch}{1.3} 
\caption{Summary of parameter priors and best-fit values for the four models.}
\vspace{-1em}
\label{tab:parameters}
\begin{tabular}{lccccccccc}
\hline
\hline
\textbf{Parameter} & \multicolumn{3}{c}{\textbf{Prior $^{\textit{a}}$ }} & & \multicolumn{4}{c}{\textbf{Best-fit Values}$^{\textit{b}}$} & \textbf{Units} \\ 
\cline{2-4} \cline{6-9}
 &  Shape & Min/$\mu$ & Max/$\sigma$ & & \citetalias{SW2017} (n=1.5) & \citetalias{SW2017} (n=3) & \citetalias{Piro2021} & \citetalias{MSW2023} &   \\ 
\hline
Progenitor Radius ($R$) & Uniform & 0 & 100 &  & $35.71^{+7.14}_{-5.71}$ & $50.0^{+14.28}_{-8.57}$ & $14.61  \pm 0.36$ & $57.14 \pm 7.20$ & $R_\odot$ \\
Envelope Mass ($M_e$) & Uniform & 0 & 10  & & $0.7^{+0.2}_{-0.3}$ & $0.8^{+0.1}_{-0.2}$ & $1.35^{+1.97}_{-0.87}$ & $0.7^{+0.2}_{-0.3}$ & $M_\odot$ \\
Shock velocity ($v_s$) & Uniform & 0 & 10 &  & $0.60^{+0.3}_{-0.2}$ & $0.67^{+0.3}_{-0.2}$   & $2.02^{+0.32}_{-0.24}$ & $0.51^{+0.4}_{-0.02}$ & $10^{4}$ km s$^{-1}$ \\
Explosion Time  ($t_{0}$) & Uniform & 60558.0 & 60560.5 & & $60558.90^{+0.07}_{-0.18}$ & $60558.80^{+0.1}_{-0.3}$& $60560.10 \pm 0.01$ &$60559.30^{+0.1}_{-0.2}$& MJD \\
Ejecta mass $\times$ factor$^{\textit{c}}$ ($f_\rho M$) & Uniform & 3 & 100 & & $50^{+30}_{-40}$ & $60 \pm 30$& -- & $60 \pm 30$ & $M_\odot$ \\
Intrinsic scatter ($\sigma$)$^{\textit{d}}$ & Half-Gaussian & 0 & 100 &  & $6.8^{+0.9}_{-0.8}$ & $7.8^{+1.2}_{-0.9}$ & -- & $3.4^{+0.5}_{-0.4}$ & -- \\
\hline
\end{tabular}
\begin{tablenotes}
\footnotesize
\item \textbf{Notes:} 
\item $^{\textit{a}}$ Prior column lists the minimum and maximum for a uniform distribution, and the mean and standard
deviation for a Gaussian distribution.
\item $^{\textit{b}}$ Best-fit values represent the 16th, 50th and 84th percentiles of the posterior distribution. \citetalias{SW2017}, \citetalias{Piro2021} and \citetalias{MSW2023} are shock cooling models detailed in \citet{SW2017}, \citet{Piro2021} and \citet{MSW2023}, respectively.
\item $^{\textit{c}}$ The ejecta mass do not have a strong effect on the early shock cooling part of the light curve. Therefore this parameter is essentially unconstrained.
\item  $^{\textit{d}}$  The \citetalias{Piro2021} model was fit using the \texttt{shock-cooling-curve} package \citep{shock_cool_curve}, while the \citetalias{SW2017} and \citetalias{MSW2023} models were employed via the Light Curve Fitting package \citep{hosseinzadeh_2024_11405219}.
\end{tablenotes}
\end{table*}

Most spherical phase models rely on polytropic envelope structures, with \citet{Piro2021}, hereafter \citetalias{Piro2021}, introducing a broken power law representation of the density profile, applicable to shock breakout conditions. Interestingly, \citet{SW2017}, hereafter \citetalias{SW2017}, demonstrated that shock cooling emission is relatively insensitive to the polytropic index and exhibits only a weak dependence on the progenitor density structure. Opacity effects, particularly those from bound-free and bound-bound transitions, play a key role in shaping the emitted radiation. While \citet{Rabinak2011} and \citet{SW2017} incorporated detailed contributions of opacity in their analysis, \citet{Nakar2010} and \citet{Shussman2016} used simplified treatments of bound-free opacity in hydrogen. The spherical phase models generally describe emission arising from the outermost layers of the envelope, with \citet{SW2017} extending these descriptions to later times by incorporating numerical simulations that account for radiation from deeper layers with complex density profiles. The transition from the planar to the spherical phase has also been explored. \citet{Shussman2016} developed an interpolation model for this transition and showed how they can be calibrated against numerical results. 

Recent refinements to shock cooling models by \citet{MSW2023}, hereafter \citetalias{MSW2023}, use similar interpolation methods while calibrating against hydrodynamic simulations covering a broad range of progenitor properties. \citetalias{MSW2023} combines solutions from \citet{SKW2011} and \citet{Katz2012} for the planar phase with those of \citet{Rabinak2011} and \citetalias{SW2017} for the spherical phase. This model is then calibrated against numerical hydrodynamic simulations spanning explosion energies of
$10^{50}$ - $10^{52}$ erg and progenitor properties such as masses of 2-40 $M_{\odot}$, radii of $3 \times 10^{12}$-$10^{14}$ cm, core to envelope mass ratios of $10^{-0.1}$ to $10^{0.1}$, and metallicities of 0.1-1 $Z_{\odot}$. These simulations assume local thermodynamic equilibrium and diffusion-based radiation transport with a constant electron scattering opacity ($0.34 \, \mathrm{cm}^2 \, \mathrm{g}^{-1}$) , which provides accurate results for highly ionized plasma ($T \geq 0.7 \, \mathrm{eV}$), as demonstrated by \citetalias{SW2017}. \citetalias{MSW2023} additionally account for line blanketing effects and have been used to model early CCSNe light curves \citep{GH2023, Irani2024, Meza_Retamal_2024, MS2024_23axu}.

We analyze the early multi-wavelength dataset of SN~2024uwq displaying shock cooling emission by fitting it to the models described in \citetalias{SW2017}, \citetalias{Piro2021}, and \citetalias{MSW2023}. The \texttt{shock-cooling-curve} package \citep{shock_cool_curve} is employed to fit the \citetalias{Piro2021} model, while the \citetalias{SW2017} and \citetalias{MSW2023} models are fit to the early light curve using a Markov Chain Monte Carlo (MCMC) routine implemented in the Light Curve Fitting package \citep{hosseinzadeh_2024_11405219}. For the \citetalias{SW2017} model, we consider two polytropic indices (\(n = 3/2\) and \(n = 3\)), corresponding to convective and radiative envelopes, respectively. The MCMC routine is utilized to fit the following parameters across all four models: progenitor radius (\(R\)), shock velocity scale (\(v_s\)), and envelope mass (\(M_{\text{env}}\)). For the model fits implemented using Light Curve Fitting, we additionally incorporate an intrinsic scatter term (\(\sigma\)) to account for scatter around the model and the potential underestimation of photometric uncertainties. The observed error bars are scaled by a factor of \(\sqrt{1 + \sigma^2}\). An additional parameter, the product of the total ejecta mass and a constant of order unity (\(f_\rho\), hereafter referred to as ``scaled ejecta mass'' \(f_\rho M\)), is included; however, the ejecta mass and density profile exhibit minimal influence on the light curve, rendering this parameter effectively unconstrained. 

To ensure the validity of the data in accordance with the models, we select observations taken within the first 3.5 days after the explosion, where the effective temperature (\(T_{\text{eff}}\)) is less than 0.7 eV, since this regime is well described by the shock cooling emission models under consideration. We performed MCMC sampling using 100 walkers initialized across the parameter space. The chains were run for 5000 steps to ensure convergence, which was assessed through visual inspection of the chain histories and by ensuring that the autocorrelation time indicated sufficient mixing of the chains. An additional 1000 steps were performed to sample the posterior distribution thoroughly. The adopted priors and derived best-fit parameter values are summarized in Table~\ref{tab:parameters}, and the resulting best-fit model is presented in Figure~\ref{fig:sc_early}. We analyze these results and discuss their implications for progenitor scenarios in the next section.

\section{Results and Discussion}\label{sec:results}

All shock-cooling models applied in this work reproduce the early observations of SN~2024uwq reasonably well, with a few notable distinctions. As seen in Figure \ref{fig:sc_early}, each model predicts an early UV/optical peak followed by a decline, which is steeper in the UV bands (e.g., \textit{W2}, \textit{M2}, \textit{W1}, \textit{U}) compared to the subtler decline in optical passbands (\textit{B}, \textit{g}, \textit{V}, \textit{r}, \textit{o}, \textit{i}). The models diverge in their ability to match the earliest UV detections and the inferred explosion epochs. The \citetalias{Piro2021} model yields an explosion epoch closest to the first ATLAS \textit{o}-band detection, whereas the \citetalias{SW2017} ($n = 1.5$ [convective] and $n = 3$ [radiative]) and \citetalias{MSW2023} models predict earlier onset of shock-cooling signatures compared to the ATLAS \textit{o}-band detection. The \citetalias{MSW2023} model provides the most consistent fit to the earliest ATLAS-\textit{o} detections in comparison with the other models. Both the  \citetalias{SW2017} models over predict fluxes in the earliest \textit{Swift} UV bands, whereas the \citetalias{Piro2021} model underestimates UV fluxes at $t > 1.5$~days. The \citetalias{MSW2023} model better matches the UV light-curve morphology at all phases, owing to its updated treatment of radiative transfer as well as line blanketing effects. The steep density gradients as assumed in \citetalias{Piro2021} favor compact progenitors (with \(R \approx 14.6\,R_\odot\)), while models with gradual density profiles (\citetalias{MSW2023}) align with more extended progenitors (up to \(R \approx 57\,R_\odot\)). The tighter constraints in \citetalias{Piro2021} arise from non-exclusion of intrinsic scatter parameter that reflect the systematic uncertainties from unmodeled envelope inhomogeneities or density gradients, whereas the other models adopts broader priors for $\sigma$, reflecting more conservative error estimates. All of the models consistently fit the optical data well, underscoring the importance of early UV observations in disentangling the progenitor structures by breaking any known degeneracies.

The inferred progenitor properties are inconsistent with those of classical red supergiants (RSGs), which typically have radii \(R \gtrsim 100\,R_\odot\) \citep[e.g.][]{GH2023, Meza_Retamal_2024, Andrews_2024, Shreshta_ggi}. The envelope mass for SN~2024uwq ($M_{\mathrm{e}} \sim 0.7$--$1.4\,M_\odot$) is larger than SN~1993J, $M_{\mathrm{e}} \sim 0.4\,M_\odot$ \citep{Woosley1994a}, which explains the persistence of weak H lines at $> 50$ days. This estimate is, however smaller than the hydrogen-rich envelopes of typical Type IIP SNe ($M_{\mathrm{e}} \gtrsim 4-10\,M_\odot$; \citep{Jerkstand2012, Dessart2013, Sukhbold2018}). This intermediate value suggests a progenitor that retained a modest hydrogen envelope prior to explosion, likely stripped via binary interactions. The results from the best-fit models therefore constrain SN~2024uwq's progenitor likely to be a stripped blue/yellow supergiant (BSG/YSG) with a radius $R = 14.6\text{--}57.1\,R_\odot$, and a hydrogen envelope mass $M_e = 0.7\text{--}1.35\,M_\odot$.  

Unlike some well-studied Type IIb SNe (e.g., SN ~1993J; \citealt{Aldering1993}, SN~2008ax; \citealt{C2008}, SN~2011dh; \citealt{Maund2011}, SN~2013df; \citealt{VanDyk2014}, SN~2016gkg; \citealt{Kilpatrick2022}, SN~2024abfo; \citealt{Regutti2025}), there is no pre-explosion imaging of the SN~2024uwq progenitor to directly constrain its pre-SN luminosity or radius. Nonetheless, the inferred properties align with the observed continuum of Type IIb progenitors, ranging from ultra-stripped systems like SN~2011ei ($M_{\mathrm{e}} < 0.1\,M_\odot$; \citealt{Dan2013}) to moderately stripped SN~1993J ($M_{\mathrm{e}} < 0.4\,M_\odot$; \citealt{Woosley1994a}), to minimally stripped cases like SN~2017jgh ($M_{\mathrm{e}} < 1\,M_\odot$; \citealt{Armstrong_2017jgh}). Standard stellar evolution models that invoke binaries predict extended RSGs for stars with $\geq 1\,M_\odot$ hydrogen at collapse, as compared to the compact BSG/YSG inferred for SN~2024uwq from analytical model fits \citep{Yoon2017}. This makes SN~2024uwq an intriguing case, potentially indicating a continuous evolutionary spectrum between canonical Type II and SESNe. This continuum is theorized to reflect the efficiency of mass transfer in binary systems, where initial orbital parameters (i.e. mass ratios or orbital periods) or less efficient stripping could result in the larger envelope mass observed here \citep{Claeys2011,Smith2011, S2020}.

\begin{figure}[tp]
   \centering
\includegraphics[width=\linewidth]{ 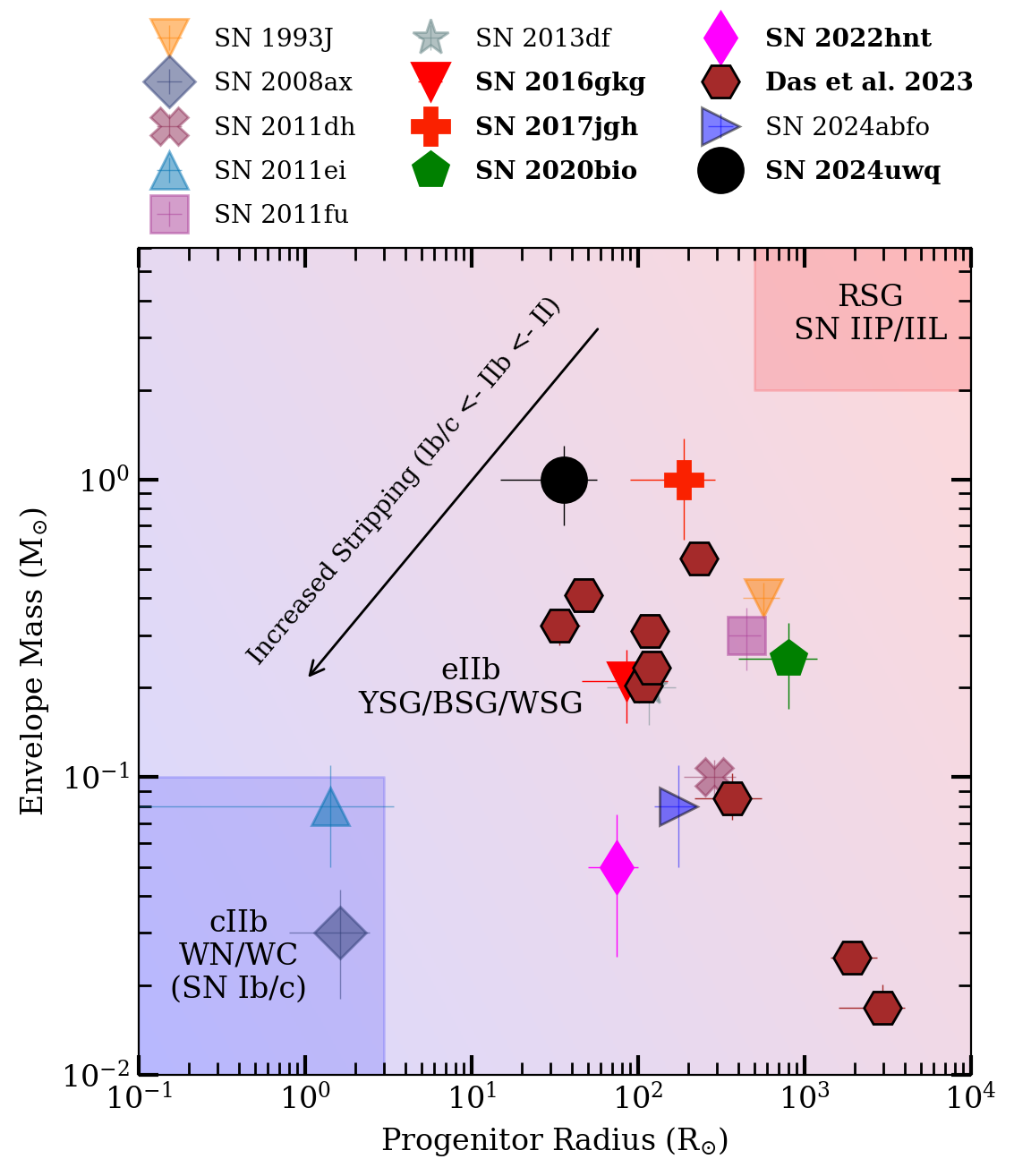}
\vspace{-2em}
    \caption{Progenitor radius vs. envelope mass for Type IIb SNe, including SN~2024uwq. Median values and 1-$\sigma$ uncertainties from the published ranges are obtained from: SN~1993J: \citealt{Woosley1994a}; SN~2008ax: \citealt{Pastorello2008,CS2010}; SN~2011dh \citealt{B2012}; SN~2011ei: \citealt{Dan2013}; SN~2011fu: \citealt{Morales2015_2011fu}; SN~2013df: \citealt{morales2014_2013df}; SN~2016gkg: \citealt{Arcavi2017}; SN~2017jgh: \citealt{Armstrong_2017jgh}; SN~2020bio: \citealt{Pellegrino_2023_SN2020bio}; SN~2022hnt: \citealt{F2025_2022hnt};
    SN~2024abfo: \citealt{Regutti2025} and
    \citealt{Das2023}.
 }
   \label{fig:phase_space}
   \vspace{-1 em}
\end{figure}

Shock velocity estimates from model fits support a YSG/BSG progenitor scenario. The high velocity obtained from the \citetalias{Piro2021} model (\(v_s = 2.02^{+0.32}_{-0.24} \times 10^{4}\,\text{km\,s}^{-1}\)) is consistent with early-time H\(\alpha\) expansion velocities (\(v_{\rm H\alpha} \sim 15{,}500\,\text{km\,s}^{-1}\)), and aligns with expectations for more compact progenitors than RSGs, where the shock propagates through a dense, steeply stratified envelope.  RSGs, in contrast, exhibit slower shock velocities (\(v_s \lesssim 10^4\)\text{km\,s}\(^{-1}\)) due to their extended, low-density envelopes \citep{Shreshta_ggi, Andrews_2024}. The lower velocities  (\(\lesssim 0.7 \times 10^{4}\,\text{km\,s}^{-1}\)) predicted by other shock cooling models are more consistent with later-phase He I measurements (\(v_{\rm He\,I} \sim 7{,}000\,\text{km\,s}^{-1}\)), which trace deeper, slower-moving ejecta. Moreover, the \citetalias{SW2017} and \citetalias{MSW2023} models yield \(v_s\) values that align with theoretical expectations \citep[\(v_s \approx 0.5\,v_{\rm exp}\);][]{Matzner1999}, where \(v_{\rm exp}\) represents the velocity of the outermost ejecta layers.

The two-zone Arnett model described in Section \ref{subsec: bol_analysis}, yields an ejecta mass of \(M_{\rm ej} \approx 3.0\,M_\odot\), and when combined with a typical neutron star remnant (\(\sim1.4\,M_\odot\)), the pre-supernova mass is $\sim$ \(4.4\,M_\odot\). Stellar evolution models indicate that single stars with initial (ZAMS) masses of \(12\text{--}20\,M_\odot\) develop helium cores of \(4\text{--}6\,M_\odot\) by core collapse \citep{Smartt2015}. This range overlaps with the inferred \(M_{\rm pre\text{-}SN}\), but the retained hydrogen envelope mass (\(M_{\rm env} \sim 0.7\text{--}1.35\,M_\odot\)) presents a challenge for single-star models. Radiation-driven winds in stars \(< 20\,M_\odot\) are inefficient, stripping \(< 0.1\,M_\odot\) of hydrogen \citep{Vink2011}, which is insufficient to account for the inferred envelope mass in SN~2024uwq. At solar or sub-solar metallicities, even enhanced single-star mass loss due to rotation (e.g., pulsational instabilities or eruptions) could only strip \(\lesssim 0.3\,M_\odot\) \citep{Yoon2017}. Given the considerations to single-star progenitors, these models require fine-tuned initial masses, metallicities and rotations, making this scenario less probable.

Binary-driven mass loss could more naturally explain the intermediate envelope mass. Case B/C mass transfer in binaries can strip hydrogen envelopes efficiently to \(M_{\rm env} \sim 0.1\text{--}1.5\,M_\odot\) \citep{Yoon2017}, matching with SN~2024uwq's properties. The derived \(M_{\rm pre\text{-}SN} \approx 4.4\,M_\odot\) aligns with stripped helium cores of \(M_{\rm ZAMS} \sim 12\text{--}15\,M_\odot\) stars that were the product of binary interaction \citep{Laplace2021, Vartanyan2021}. While direct evidence for a binary companion is absent in the existing data, the intermediate \(M_{\rm env}\) inferred from the shock cooling models favors binary stripping over single-star winds. Wolf-Rayet progenitors, which lose nearly their entire hydrogen envelopes (\(M_{\rm env} < 0.1\,M_\odot\)) via strong winds are unlikely as these systems  typically produce more massive helium cores (\(M_{\rm He} > 6\,M_\odot\)) and higher explosion energies \citep{Sukhbold2016}, inconsistent with the inferred \(M_{\rm ej}\) and shock velocities for SN~2024uwq.

We place SN~2024uwq in context with other Type IIb supernovae in the $R$--$M_{\rm env}$ phase space shown in Figure~\ref{fig:phase_space}.  SNe highlighted in bold have progenitor radius and mass estimates derived solely from early shock-cooling analysis; the rest are based on alternative methods that includes pre-explosion imaging, late-time radio observations and hydrodynamical modeling. \citet{CS2010} proposed that Type IIb SNe may be divided into two subgroups: extended-envelope IIb (eIIb; e.g., SN~1993J, with $M_{\rm env} > 0.1\,M_\odot$, $R_{\rm env} \sim 10^{13}$\,cm) and compact-envelope IIb (cIIb; e.g., SN~2008ax, with $M_{\rm env} < 0.1\,M_\odot$, $R_{\rm env} \sim 10^{11}$\,cm). These subtypes may represent a continuum modulated by the amount of residual hydrogen, with cIIb events potentially bridging to Type Ib SNe.
SN~2024uwq, with intermediate progenitor properties ($M_{\rm env} \sim 0.7$--$1.35\,M_\odot$, $R_{\rm env} \sim 14.6$--$57.1\,R_\odot$), occupies a transitional region in this space, possibly resulting from moderate stripping in a binary system. 

This continuum, spanning objects like SN~2008ax (cIIb) and SN~2017jgh (eIIb), underscores varying degrees of envelope stripping, likely modulated by binary interaction efficiency. Detailed hydrodynamical modeling of both single stars with mass loss and interacting binary systems, combined with comprehensive observational studies, is essential to advance our understanding of the complex evolutionary pathways of massive stars and their role in shaping the diversity of Type~IIb supernovae \citep{Long2022, Goldberg2022, Hayne2023}. Our findings in this study highlight the necessity of early UV observations to resolve shock-cooling phases, which are pivotal for constraining progenitor radii and mass-loss histories of massive stars.

\subsection{Future Surveys and SN IIb rates}

\begin{figure}[tp]
   \centering
\includegraphics[width=\linewidth]
{ 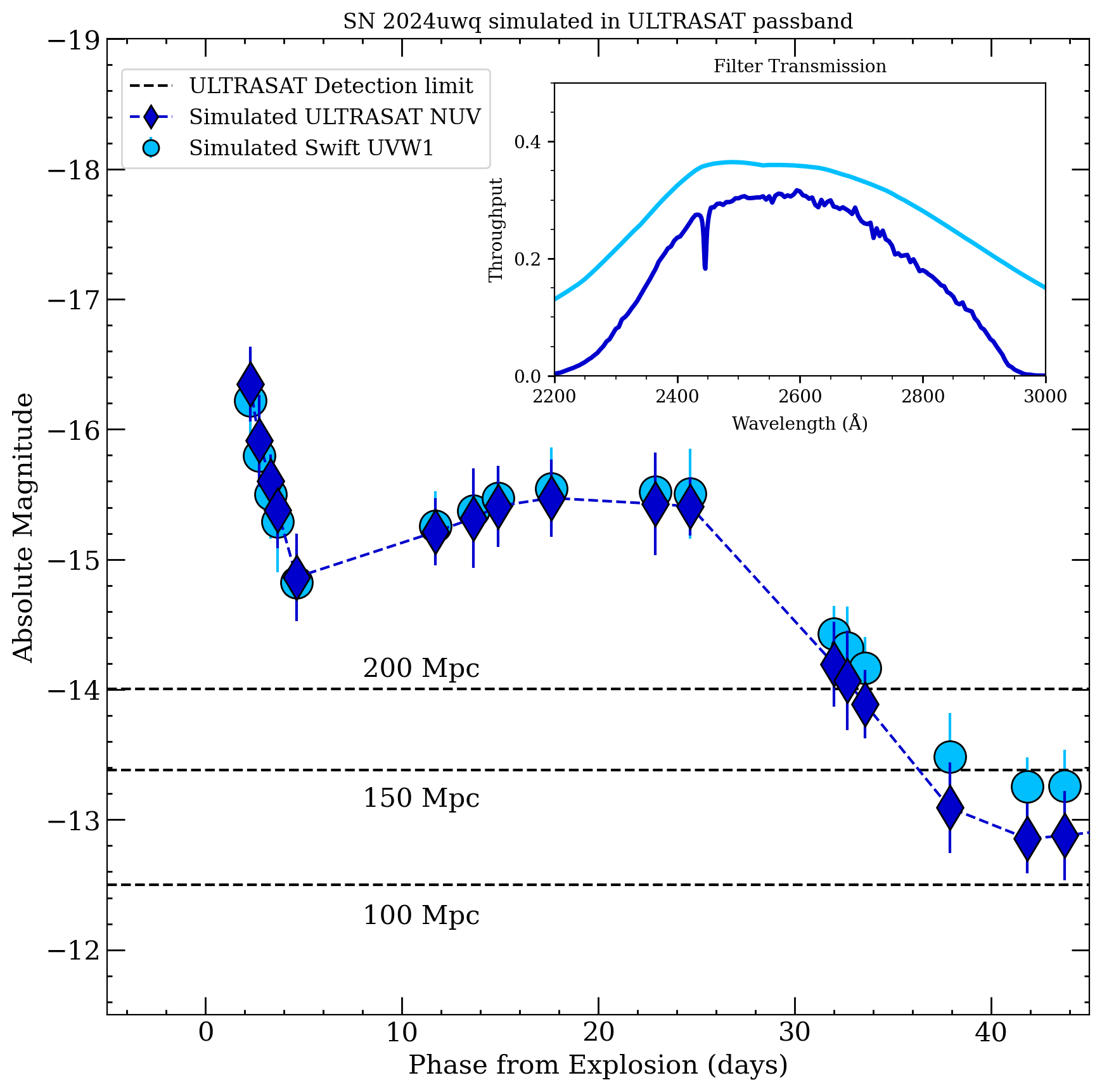}
\vspace{-1 em}
\caption{Simulated \textit{ULTRASAT} NUV and \textit{Swift} UVW1 light curves, demonstrating \textit{ULTRASAT}'s detectability of complete shock cooling emission phases out to 200 Mpc. Photometry was simulated using the SED of SN~2024uwq at each epoch convolved with the respective filter throughput.
The error bars represent the photometric uncertainties. }
\label{fig:uwq_RMC_ULTRASAT}
\vspace{-1 em}
\end{figure}

Future time-domain surveys, including \textit{ULTRASAT}, \textit{UVEX}, and the Vera Rubin Observatory's LSST, will revolutionize our understanding of stripped-envelope SNe progenitors and massive star evolution \citep{Ivezic2019, Kulkarni2021, Yossi2024}. Figure \ref{fig:uwq_RMC_ULTRASAT} shows the simulated NUV light curves for SN~2024uwq by convolving its observed spectral energy distribution at each epoch with the \textit{ULTRASAT} and \textit{Swift} UVW1 filter throughputs, demonstrating its ability to capture the complete shock-cooling emission phase out to ~200 Mpc. \textit{ULTRASAT}'s wide-field UV coverage will detect shock-cooling emission within hours of an explosion, resolving the blue excess of the shock breakout phase that optical surveys currently miss. This will deliver high-cadence, high-quality NUV light curves for these early stages. 

Using CCSN rates derived from \citet{Li2011} to be \(7.05 \pm 1.56 \times 10^{-5}\) SN Mpc\(^{-3}\) yr\(^{-1}\)and assuming Type IIb constitute 10.6\% of CCSNe \citep{Smith2011}, the estimate for local Type IIb rate is approximately \(0.747 \times 10^{-5}\) SN Mpc\(^{-3}\) yr\(^{-1}\). Assuming an initial peak of M$_{NUV}$ $\sim$ \(-17\) for Type IIb, \textit{ULTRASAT}'s 200\,deg\(^2\) field of view and 22.5\,mag sensitivity \citep{Yossi2024}, yield a detection rate of $\sim$ 77-100 SN IIb per year, for a detection horizon upto $\sim$800 Mpc. However, for a SN~2024uwq analogous Type IIb, only those within $\lesssim$200 Mpc will have well sampled light curves ($\leq 20$ days) to fully characterize the shock-cooling phase, as shown in Figure \ref{fig:uwq_RMC_ULTRASAT}. This reduced detection horizon, intrinsic diversity in initial SESNe peak UV luminosities ($M_{\mathrm{NUV}} \sim -16$ to $-18$) and extinction considerations, combined with SN IIb volumetric rates and ULTRASAT's field of view, suggests that only $\sim 2-3$ SN IIb events per year will be observed for detailed progenitor studies. \textit{UVEX} will complement \textit{ULTRASAT} by providing synoptic UV spectroscopy, critical for disentangling line-forming regions in early phases, while LSST's deep optical cadence, though less frequent in the UV, will deliver statistically robust populations to contextualize binary fractions and explosion asymmetries.

\section{Conclusions}\label{sec:conclusion}

We have presented early high-cadence multiwavelength photometric and spectroscopic observations of SN~2024uwq, a Type IIb SN with characteristic shock cooling emission. Our main conclusions are summarized below.

\begin{enumerate}
    \item The early light curve of SN~2024uwq exhibits a double-peaked profile, with an initial maximum of $M_B = -16.3\,\text{mag}$ at $t \approx 2\,\text{days}$, followed by a brighter second peak at $M_B = -17.5\,\text{mag}$ at $t \approx 20\,\text{days}$. The shock-cooling phase lasts $\sim 5$ days, suggesting a partially stripped progenitor.

    \item The early phase spectra display broad H\(\alpha\) and He I P-Cygni profiles with initial velocities of \(\sim15\,500\) km s\(^{-1}\) and \(\sim14\,000\) km s\(^{-1}\) that decline to \(\sim10\,000\) km s\(^{-1}\) and \(\sim7\,000\) km s\(^{-1}\) by \(t \approx 30\,\text{days}\), reflecting a hydrogen envelope overlying the He rich ejecta.

    \item The Arnett-model fits yield ejecta and nickel parameters of \(M_{\rm ej} = 3.00^{+0.10}_{-0.09}\,M_\odot\), \(E_k = 2.75 \times 10^{51}\,\text{erg}\), and \(M_{\rm 56Ni} = 0.098\,M_\odot\), which are consistent with typical stripped-envelope SNe IIb.

    \item Shock cooling models constrain the progenitor's radius to \(R = 14.6\text{--}57.1\,R_\odot\) and the hydrogen envelope mass to \(M_e = 0.7\text{--}1.35\,M_\odot\). The observed shock velocities (\(v_s \approx 0.5\text{--}2.0 \times 10^4\,\text{km\,s}^{-1}\)) agree with theoretical expectations.
    
    \item The progenitor was likely a \(12\text{--}20\,M_\odot\) ZAMS star that evolved to become a BSG/YSG likely in a binary system, positioning SN~2024uwq within the observed continuum of Type IIb SNe, ranging from highly stripped events (e.g., SN~2008ax, SN~2011ei) to minimally stripped ones (e.g., SN~1993J, SN~2017jgh).
\end{enumerate}

The synergy of \textit{ULTRASAT}'s wide-field sky coverage, \textit{UVEX}'s spectroscopic characterization, and LSST's comprehensive volumetric surveys promises to disentangle current model degeneracies in stripped-envelope supernova progenitor scenarios and mass-loss mechanisms. This integrated approach will move beyond the study of individual transients like SN~2024uwq, enabling population-level analyses of rich multi-wavelength datasets that will provide critical insights into the evolutionary pathways of massive stars. 

\section*{acknowledgments}

We thank Yossi Shvartzvald for generously providing the total throughput, limiting magnitude, and source data for \textit{ULTRASAT}, used in this work. Time domain research by the University of Arizona team and D.J.S. is supported by National Science Foundation (NSF) grants 2108032, 2308181, 2407566, and 2432036 and the Heising-Simons Foundation under grant $\#$2020-1864. This work makes use of data from the Las Cumbres Observatory global telescope network, which is supported by NSF grants AST-1911225 and AST-1911151. S.V. and the UC Davis time-domain research team acknowledge support by NSF grants AST-2407565. J.E.A. is supported by the international Gemini Observatory, a program of NSF's NOIRLab, which is managed by the Association of Universities for Research in Astronomy (AURA) under a cooperative agreement with the National Science Foundation, on behalf of the Gemini partnership of Argentina, Brazil, Canada, Chile, the Republic of Korea, and the United States of America.  K.A.B. is supported by an LSST-DA Catalyst Fellowship; this publication was thus made possible through the support of Grant 62192 from the John Templeton Foundation to LSST-DA. Seong Hyun Park and Sung-Chul Yoon are supported by the National Research Foundation of Korea (NRF RS-2024-00356267).

The SALT spectra presented here were obtained through the Rutgers University SALT programs 2023-1-MLT-008, 2023-2-SCI-030 and 2024-1-MLT-003 (PI: Jah). Observations for this work were also obtained at the Southern Astrophysical Research (SOAR) telescope, which is a joint project of the Minist\'{e}rio da Ci\^{e}ncia, Tecnologia e Inova\c{c}\~{o}es (MCTI/LNA) do Brasil, the US National Science Foundation's NOIRLab, the University of North Carolina at Chapel Hill (UNC), and Michigan State University (MSU). We acknowledge the important contribution of David Sanmartim, who developed the initial incarnation of the redccd module. The authors thank Tina Armond for her invaluable help in adding calibrated comparison lamps to the library of reference comparison lamps for wavelength solution. Our work would not be possible without the friendly work atmosphere at CTIO headquarters in La Serena, where we can interact with our SOAR and CTIO colleagues in lively and useful discussions that have been important in making the Goodman pipeline possible. We also acknowledge fruitful discussions and suggestions from our colleagues Bart Dunlop, Chris Clemens, and Erik Dennihy at the University of North Carolina at Chapel Hill.

This work has made use of data from the Asteroid Terrestrial-impact Last Alert System (ATLAS) project. The Asteroid Terrestrial-impact Last Alert System (ATLAS) project is primarily funded to search for near-Earth asteroids through NASA grants NN12AR55G, 80NSSC18K0284, and 80NSSC18K1575; byproducts of the NEO search include images and catalogs from the survey area. This work was partially funded by the Kepler / K2 grant J1944/80NSSC19K0112, HST GO-15889, and the STFC grants ST/T000198/1 and ST/S006109/1. The ATLAS science products were made possible through the contributions of the University of Hawaii Institute for Astronomy, Queen's University Belfast, the Space Telescope Science Institute, the South African Astronomical Observatory, and The Millennium Institute of Astrophysics (MAS), Chile. This research has made use of the NASA Astrophysics Data System (ADS) Bibliographic Services and the NASA/IPAC Infrared Science Archive (IRSA), which is funded by the National Aeronautics and Space Administration and operated by the California Institute of Technology. This work also made use of data supplied by the UK Swift Science Data Centre at the University of Leicester.

\facilities{ADS, Neil Gehrels \textit{Swift} Observatory (UVOT), AAVSO, NED,  Las
Cumbres Observatory (Sinistro, FLOYDS), Southern Astrophysical Research
Telescope (SOAR: Goodman), SALT (RSS), Meckering: PROMPT, ATLAS, WISeREP}

\software{Astropy \citep{astropy:2013,astropy:2018, astropy:2022}, Photutils \citep{Bradley_2022}, BANZAI \citep{Banzai}, Light Curve Fitting \citep{hosseinzadeh_2024_11405219}, \texttt{emcee} \citep{FM2013_emcee}, \texttt{corner} \citep{FM2016_corner}, \texttt{Superbol} \citep{Nicoll2018},  \texttt{shock-cooling-curve} \citep{shock_cool_curve}, Matplotlib \citep{mpl}, Numpy \citep{numpy}, Scipy \citep{scipy}, IRAF \citep{iraf1,iraf2}, PySALT \citep{PySALT}, Goodman-HTS pipeline \citep{Goodman-HTS}, FLOYDS pipeline \citep{Valenti_2014}, \texttt{lcogtsnpipe}\citep{Valenti_2016}}

\appendix

\begin{table*}[ht]
 \caption{Log of Spectroscopic Observations}
 \setlength{\tabcolsep}{12pt}
 \begin{tabular}{ c c c c c c}
    \hline
    Date (UTC) & MJD &  Telescope & Instrument & Phase (d)  & Exp (s) \\
    \hline
    2024-09-09 13:06:50 & 60562.54 & COJ   & FLOYDS &  3.91 & 1800 \\
    2024-09-09 22:50:20 & 60562.95 & SALT & RSS & 4.31 & 1133 \\
    2024-09-10 10:02:18 & 60563.41 & COJ   & FLOYDS &  4.78 & 1800 \\
    2024-09-13 11:58:50 & 60566.49 & COJ   & FLOYDS &  7.86 & 1800 \\
    2024-09-15 13:24:33 & 60568.55 & COJ   & FLOYDS &  9.92 & 1800 \\
    2024-09-16 21:59:53 & 60569.91 & SALT & RSS & 11.28 & 1500 \\
    2024-09-18 10:39:17 & 60571.44 & COJ   & FLOYDS &  12.81 & 1800 \\
    2024-09-22 11:28:15 & 60575.47 & COJ   & FLOYDS &  16.84 & 1800 \\
    2024-09-28 02:05:25 & 60581.08 & SOAR & GHTS-Red & 22.45 & 400 \\ 
    2024-09-28 10:18:09 & 60581.42 & COJ   & FLOYDS &  22.79 & 1800 \\
    2024-10-01 10:19:49 & 60584.43 & COJ   & FLOYDS &  25.79 & 1800 \\
    2024-10-03 21:08:56 & 60586.88 & SALT & RSS & 28.2 &  1500 \\
    2024-10-05 03:48:50 & 60588.15 & SOAR & GHTS-Red & 29.52 & 400 \\ 
    2024-10-07 20:37:40 & 60590.85 & SALT & RSS & 32.22 & 1500  \\
    2024-10-10 12:12:34 & 60593.50 & COJ   & FLOYDS &  34.87 & 1800 \\
    2024-10-17 19:57:33 & 60600.83 & SALT & RSS & 42.19 &  1500 \\
    2024-10-19 15:58:16 & 60602.66 & SALT & RSS & 44.03 & 1500 \\
    2024-10-20 11:35:34 & 60603.48 & COJ   & FLOYDS &  44.84 & 1800 \\
    2024-10-22 02:57:21 & 60605.12 & SOAR & GHTS-Red & 46.48 & 400 \\
    2024-10-25 11:29:18 & 60608.47 & COJ   & FLOYDS & 49.84  & 1800 \\
    2024-10-29 11:08:12 & 60612.46 & COJ   & FLOYDS &  53.83 & 1800 \\
    2024-11-03 10:12:11 & 60617.42 & COJ   & FLOYDS &  58.79 & 1800 \\
    2024-11-08 10:35:24 & 60622.44 & COJ   & FLOYDS &  63.80 & 1800 \\
    2024-11-18 10:35:24 & 60632.50 & Gemini-S   & FLAMINGOS-2 &  76.40 & 1800 \\

    \hline
 \end{tabular}
 
 \label{tab:specInst}
\end{table*}

\vspace{-3em}

\bibliography{references}{}
\bibliographystyle{aasjournal}

\end{document}